\documentclass[letterpaper,twocolumn,10pt]{article}
\usepackage{usenix-2020-09}

\usepackage{latexsym}      
\usepackage{mathtools}	   
\usepackage{xspace}   
\usepackage{booktabs}
\usepackage{epsfig}
\usepackage{amstext}
\usepackage[dvipsnames]{xcolor}   
\usepackage{tabularx} 
\usepackage{pifont}
\usepackage[font=small,labelfont=bf]{caption}
\usepackage{balance}
\usepackage{times}
\usepackage{multirow}
\usepackage{graphicx}
\usepackage{subcaption}
\usepackage{enumitem}
\usepackage{comment}
\usepackage{algorithm}
\usepackage{algpseudocode}
\usepackage{listings}
\usepackage{ulem}
\usepackage{enumitem}

\usepackage{csvsimple}
\usepackage{siunitx}
\usepackage[utf8]{inputenc}


\renewcommand{\paragraph}[1]{\vspace{5pt}\noindent\textbf{#1:}}

\begin{document}
\pagestyle{plain}

\date{}
\title{Hopper: Modeling and Detecting Lateral Movement}

\newcommand{\authspace}{\hspace{3mm}}
\newcommand{\titlespace}{\hspace{3mm}}

\def\icsi{\raisebox{6pt}{\small $\oint$}}
\def\ucb{\raisebox{6pt}{\small $\dagger$}}
\def\ucsd{\raisebox{6pt}{\small $\star$}}
\def\dbx{\raisebox{6pt}{\small $\circ$}}
\def\figma{\raisebox{6pt}{\small $\psi$}}

\author{
    Grant Ho\ucsd\ucb\dbx\authspace
    Mayank Dhiman\dbx\authspace
    Devdatta Akhawe\figma\authspace
    \\
    Vern Paxson\ucb\icsi\authspace
    Stefan Savage\ucsd\authspace
    Geoffrey M. Voelker\ucsd\authspace
    David Wagner\ucb\authspace
\vspace{.2cm}\\
  {\small\dbx{Dropbox}}\titlespace
  {\small\ucb{UC Berkeley}}\titlespace
  {\small\ucsd{UC San Diego}}\titlespace
  {\small\figma{Figma, Inc.}}\titlespace
  {\small\icsi{International Computer Science Institute}}
}

\newcommand{\dropbox}{{Dropbox}\xspace}
\newcommand{\detector}{{Hopper}\xspace}
\newcommand{\ccsDetector}{{SAL}\xspace}

\newcommand{\ttext}[1]{{\sf #1}}

\newcommand{\eg}{e.g.,\xspace}
\newcommand{\ie}{i.e.,\xspace}


\newcommand{\startMachine}{$A$\xspace}
\newcommand{\footholdMachine}{\startMachine}
\newcommand{\targetMachine}{$Z$\xspace}
\newcommand{\budgetSym}{$B$\xspace}

\newcommand{\numScenarioClearSwitch}{1\xspace}
\newcommand{\numScenarioUnclearSwitch}{2\xspace}
\newcommand{\numScenarioIntoClient}{3\xspace}

\newcommand{\attacker}{Mallory\xspace}
\newcommand{\causalUser}{\text{Alice}\xspace}
\newcommand{\switchingLoginUser}{Bob\xspace}
\newcommand{\switchingSrc}{$Y$\xspace}
\newcommand{\switchingDest}{$Z$\xspace}
\newcommand{\maxSessionHours}{24\xspace}
\newcommand{\targetUser}{\text{TargetUser}}

\newcommand{\timeCurrent}{$t_i$\xspace}
\newcommand{\srcCurrent}{$Y$\xspace}
\newcommand{\dstCurrent}{$Z$\xspace}
\newcommand{\userCurrent}{\switchingLoginUser}

\newcommand{\loginCurrent}{$L_i$\xspace}
\newcommand{\newLogin}{\loginCurrent} 
\newcommand{\otherLogin}{$L_k$\xspace} 

\newcommand{\pathCurrent}{$P$\xspace}
\newcommand{\pathStart}{$A$\xspace}
\newcommand{\switchingServer}{$Y$\xspace}
\newcommand{\pathEnd}{$Z$\xspace}
\newcommand{\pathTargetUsers}{[$\targetUser_1$, $\targetUser_2$, ... $\targetUser_N$]\xspace}
\newcommand{\pathDestList}{$\text{Dest}_P$\xspace}
\newcommand{\allUsers}{AllUsers$_P$\xspace}
\newcommand{\focalLogin}{$L_f$\xspace}
\newcommand{\continuitySym}{$Continuity_P$\xspace}

\newcommand{\changepointLogin}{$L_c$\xspace} 
\newcommand{\switchingLoginSym}{$L_i$\xspace} 
\newcommand{\inboundSwitchingLoginSym}{$L_{i-1}$\xspace} 

\newcommand{\attackExploratory}{Exploratory Attack\xspace}
\newcommand{\attackAggressive}{Aggressive Spread\xspace}
\newcommand{\attackTargeted}{Targeted Attack\xspace}

\newcommand{\dateStart}{Jan 1, 2019\xspace}
\newcommand{\dateEnd}{Apr 1, 2020\xspace}
\newcommand{\totalMonths}{15\xspace}

\newcommand{\numFinalLoginsApprox}{3.5 million\xspace}
\newcommand{\numFinalLogins}{3,527,844\xspace}
\newcommand{\medianDailyLogins}{4,098\xspace}
\newcommand{\reductionFactorWindows}{40$\times$\xspace}
\newcommand{\numTotalLoginsAfterWindows}{19.5 million\xspace}

\newcommand{\numRawLoginsApprox}{780 million\xspace}
\newcommand{\numRawLogins}{784,459,506\xspace}

\newcommand{\numUsers}{634\xspace}
\newcommand{\numMachines}{2,327\xspace}

\newcommand{\numAutomationEdges}{30\xspace}
\newcommand{\numAutomationLoginsApprox}{16 million\xspace}

\newcommand{\numTotalBenignPruningLogins}{170,000\xspace}  
\newcommand{\numBastionPruningLogins}{2,000\xspace}

\newcommand{\numNewEdgeAlertsApprox}{24,000\xspace}

\newcommand{\dateEvalStart}{Mar 1, 2019\xspace}
\newcommand{\numEvalMonths}{13\xspace}
\newcommand{\numEvalDays}{396\xspace}  
\newcommand{\numFinalEvalLogins}{2,941,173\xspace}  
\newcommand{\numFinalEvalLoginsApprox}{2.94M\xspace}  
\newcommand{\numEvalLogins}{713,617,425\xspace}
\newcommand{\numEvalLoginsApprox}{713M\xspace}

\newcommand{\trainingWindowDays}{30\xspace}


\newcommand{\totalAttackUsersLM}{50\xspace} 
\newcommand{\numAttackUsersSuccessfulLM}{41\xspace} 
\newcommand{\numAttackUsersNoLM}{9\xspace}

\newcommand{\numAttackSynthetic}{326\xspace}  
\newcommand{\numAttackRedTeam}{1\xspace}
\newcommand{\numTotalAttacks}{327\xspace}  

\newcommand{\initialAlertBudget}{5\xspace} 
\newcommand{\numAvgDailyAlerts}{9\xspace}

\newcommand{\numAlertsClearSwitchingFromClient}{1,326\xspace}
\newcommand{\numAlertsClearSwitching}{2,216\xspace}

\newcommand{\numAlertsUnclearSwitching}{1,344\xspace}

\newcommand{\numWatchlistAlerts}{49\xspace}
\newcommand{\numTotalAlerts}{3,560\xspace}

\newcommand{\fpr}{0.0012\xspace}

\newcommand{\numDetectedClearSwitch}{138\xspace}
\newcommand{\numAttackClearSwitch}{147\xspace}
\newcommand{\numDetectedClearSwitchOneHop}{95\xspace}
\newcommand{\numDetectedClearSwitchTwoHop}{55\xspace}
\newcommand{\numDetectedClearSwitchBoth}{12\xspace}

\newcommand{\numDetectedUnclearSwitch}{171\xspace}
\newcommand{\numAttackUnclearSwitchSynthetic}{180\xspace}
\newcommand{\numUniqueDailyPaths}{2,419\xspace}

\newcommand{\tprLM}{94.5\%\xspace}
\newcommand{\numDetectedTotal}{309\xspace}
\newcommand{\detectionRate}{\numDetectedTotal~/ \numTotalAttacks\xspace}
\newcommand{\numHalfAttacks}{156\xspace}
\newcommand{\detectionRateHalf}{\numHalfAttacks~/ \numTotalAttacks\xspace}

\newcommand{\ccsBaselineTuningMaxPercent}{33\%\xspace}
\newcommand{\ccsBaselineNumAlertsHalf}{3,556\xspace}
\newcommand{\ccsBaselineNumAlertsFull}{27,927\xspace}

\newcommand{\fpReductionPrecise}{7.8$\times$\xspace}  
\newcommand{\fpReduction}{8$\times$\xspace}  
\newcommand{\improvementFactor}{\fpReduction}

\newcommand{\numFNLM}{18\xspace}
\newcommand{\numFNBrokenEnrichmentLM}{9\xspace}
\newcommand{\numFNLowBudgetLM}{9\xspace}

\newcommand{\numMinDailyAlerts}{3\xspace}
\newcommand{\numUnclearSwitchAttacksTopRank}{9\xspace}
\newcommand{\numUnclearSwitchAttacksSecondRank}{1\xspace}
\newcommand{\numUnclearSwitchAttacksThirdRank}{2\xspace}

\maketitle

\begin{abstract}
In successful enterprise attacks, adversaries often need to gain access to additional machines beyond their initial point of compromise,
a set of internal movements known as \textit{lateral movement}.
We present \detector, a system for detecting lateral movement based on commonly available enterprise logs.
\detector constructs a graph of login activity among internal machines and then identifies suspicious sequences of logins that correspond to lateral movement.
To understand the larger context of each login,
\detector employs an inference algorithm to identify the broader path(s) of movement that each login belongs to and the \textit{causal} user responsible for performing a path's logins.
\detector then leverages this path inference algorithm, in conjunction with
a set of detection rules and
a new anomaly scoring algorithm,
to surface the login paths most likely to reflect lateral movement.
On a \totalMonths-month enterprise dataset 
consisting of over \numRawLoginsApprox internal logins, 
\detector achieves a \tprLM detection rate across over 300 realistic attack scenarios, 
including one red team attack,
while generating an average of $<$~\numAvgDailyAlerts alerts per day.
In contrast, to detect the same number of attacks,
prior state-of-the-art systems would need to generate nearly \fpReduction as many false positives.
\end{abstract}

\section{Introduction}

Organizations routinely fall victim to sophisticated attacks,
resulting in billions of dollars in financial harm, the theft of sensitive data, and the disruption of critical infrastructure~\cite{breachCost,targetbreach,anthembreach,opmbreach,doebreach}. 
In many of these attacks, adversaries need to move beyond their initial point of compromise to achieve their goal~\cite{humanOperatedLM1,opmbreach,ukraineLM}.
For example, an employee compromised by a spearphishing attack often does not have all of an organization's sensitive secrets readily accessible from their machine;
thus, attackers will need to move to other machines to access their desired data.
This set of malicious \textit{internal} movements is known as \textit{lateral movement}~\cite{crowdstrikelateralmovement,trendmicrolateralmovement}.

In this work, we focus on detecting lateral movement in enterprise networks.
We present \detector, a system that uses commonly-collected log data to detect lateral movement attacks with a manageable rate of false alarms. 
\detector builds a graph of user movement (logins) between internal machines and then identifies suspicious movement paths within this graph.
While prior work has proposed similar graphical models,
these approaches have either relied on narrowly crafted signatures~\cite{liu2018latte},
leaving them unable to detect many lateral movement attacks,
or applied standard anomaly detection methods that alert on rare login paths~\cite{ccs2017lmdetecting, kent2015authentication, liu_log2vec_2019}.
Unfortunately, the scale of modern enterprises inherently produces large numbers of anomalous-but-benign logins, causing traditional anomaly detection to generate too many false alarms.

\detector overcomes these challenges by employing a different approach, which we call specification-based anomaly detection.
Our approach leverages an attack specification that captures fundamental characteristics of lateral movement as a set of key path properties (\S~\ref{sec:overview}).
This specification states that successful lateral movement attacks
will (1) switch to a new set of credentials and (2) eventually access a server that the
original actor could not access.
We then combine this specification with anomaly detection, to reduce false positives and imprecision due to the limitations of real-world data.

Our attack specification capitalizes on a key observation:
adversaries generally perform lateral movement to access a machine that their initial victim lacked access to.
Thus, as part of their lateral movement activity, attackers will need to acquire and switch to a new set of credentials that enables their sought-for access.
As a result, lateral movement paths will exhibit the two key attack properties identified in our specification.
In the context of an attack's full lifecycle, our specification observes that standard authentication logs not only provide a window into lateral movement activity, but also contain implicit artifacts of other key stages in an enterprise attack.
For example, attackers use a variety of techniques to acquire privileged credentials (as detailed in the \textit{Credential Access} and \textit{Privilege Escalation} stages of the MITRE ATT\&CK Framework~\cite{mitreattack}).
While prior work detects these other attack stages through intricate host-activity analysis~\cite{milajerdi2019holmes,hassan2020tactical,hossain2020combating},
the fruits of these malicious actions manifest themselves during lateral movement, since attackers use these new credentials to access new data and machines.
Through the detection methods that we develop,
\detector infers and leverages such signals (reflected in our two key attack properties) to help uncover lateral movement activity.

To identify paths with the two key properties, we develop methods for reconstructing a user's global movement activity from the point-wise login events reported in common authentication logs.
These methods allow \detector to infer the \textit{causal} user responsible for performing each login and the broader path of movement a login belongs to (\S~\ref{sec:paths}).
Unfortunately, real-world authentication logs do not always contain sufficient information for \detector to clearly identify the causal user who made each login,
resulting in uncertainty about whether some paths truly exhibit the two key attack properties.
To resolve these cases of uncertainty,
\detector employs a new anomaly detection algorithm to identify the most suspicious paths to alert on (\S~\ref{sec:detector}).
This selective approach to anomaly detection
is a key distinction that allows \detector to significantly outperform prior work that relies on traditional anomaly detection~\cite{ccs2017lmdetecting} or signature-based detection~\cite{liu2018latte}.

We evaluate \detector on a \totalMonths-month enterprise data set that contains over \numRawLoginsApprox internal login events (\S~\ref{sec:evaluation}).
This data includes one lateral movement attack performed by a professional red team and \numAttackSynthetic simulated attacks that span a diverse array of real-world scenarios (ranging from ransomware to stealthy, targeted machine compromise).
On this data set, \detector can detect \numDetectedTotal~/~\numTotalAttacks attacks while
generating $<$~\numAvgDailyAlerts false positives per day on average, which is
an \fpReduction improvement over prior state-of-the-art systems~\cite{ccs2017lmdetecting}.

In summary, we make the following contributions:
\begin{itemize}[noitemsep,topsep=0pt]
    \item We present \detector, a novel system that uses commonly-collected authentication logs to detect lateral movement.
    \detector employs a new detection approach based on a principled set of properties that successful lateral movement paths will exhibit (\S~\ref{sec:overview}).
    \item Our approach identifies paths with these key properties by inferring the broader paths of movement that users make (\S~\ref{sec:paths}), and strategically applies a new anomaly scoring algorithm to handle uncertainty that arises due to the limited information in real-world logs (\S~\ref{sec:detector}).
    \item We evaluate \detector on 15 months of enterprise data, including a red team attack and over 300 realistic attack simulations.
   \detector detects \tprLM of these attacks, and produces \fpReduction fewer false alarms than prior work (\S~\ref{sec:evaluation}).
\end{itemize}

\begin{figure}[t]
\centering
\includegraphics[width=0.48\textwidth]{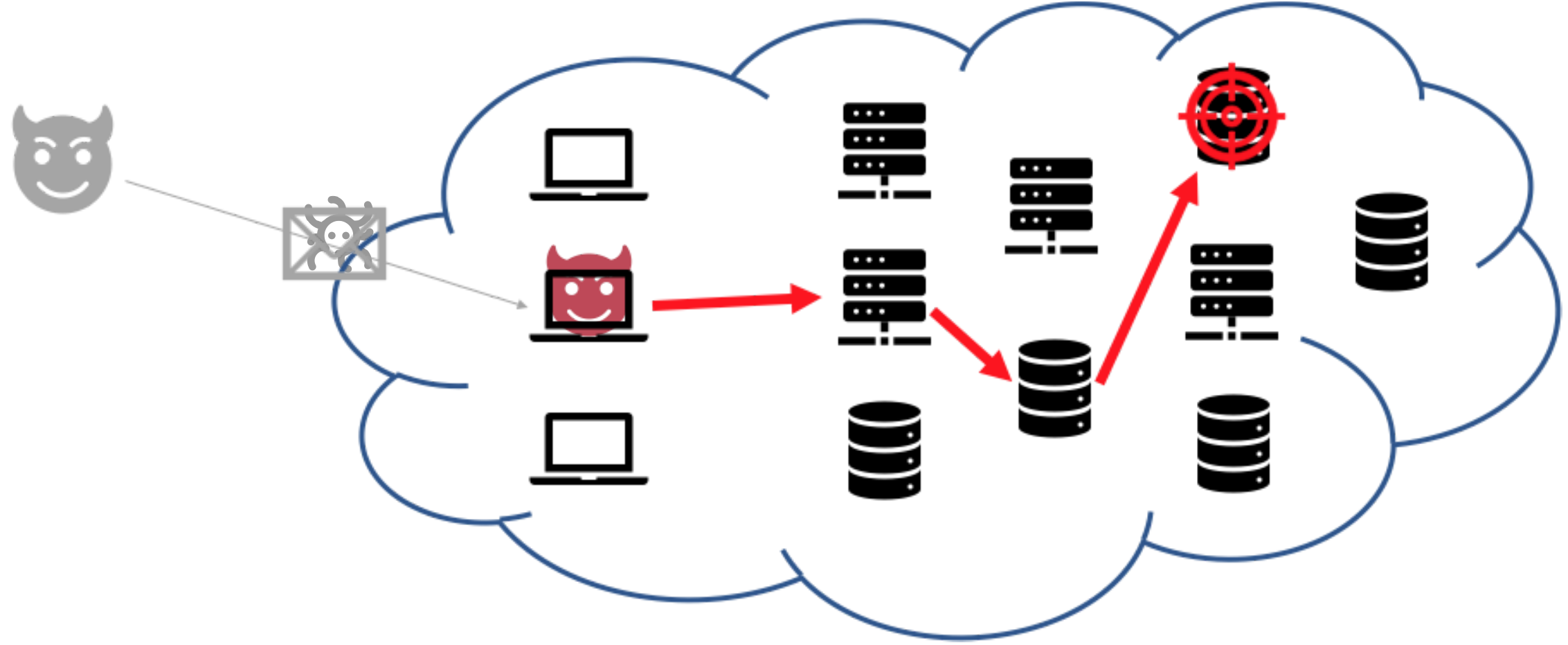}
\caption{Lateral movement, depicted as red arrows, is the set of attacker movements between \textit{internal} machines in an enterprise. 
}
\label{fig:lateral_movement}
\end{figure}

\section{Background}
\label{sec:background}
The internal movements that attackers make between machines within an enterprise is known as \textit{lateral movement} (Figure~\ref{fig:lateral_movement}).
In this section, we review prior work on defending against lateral movement and describe the goals and assumptions that underlie our detection approach.
\subsection{Related Work}
\label{sec:related}

Prior work pursues three general strategies for mitigating lateral movement:
improving security policies to limit attacker movement;
detecting lateral movement activity;
and developing forensic techniques to help remediate a known attack.
We consider the first and last lines of work as complementary directions to our work; we focus on developing practical detection for lateral movement attacks.
The first direction, proactively improving security policies, enables an organization to implement better least privilege policies and identify high-risk machines that warrant additional monitoring~\cite{dunagan2009heat, freitas2020d2m, hagberg_connected_2014, bloodhound}.
While beneficial, these policies do not fully eliminate all possible lateral movement paths;
indeed, our work aims to detect attacks that can succeed even at organizations with good least privilege hygiene.
The third line of related work, investigating a known attack, assumes that an organization has already identified the existence of a breach.
Enterprises can use these prior methods to effectively analyze and remediate a lateral movement attack identified by \detector.

Prior work on detecting lateral movement frequently models internal logins as a graph of machine-to-machine movement~\cite{liu2018latte, ccs2017lmdetecting, kent2015authentication, bohara2017unsupervised, purvine2016graph, liu_log2vec_2019, wilkens_towards_2019, lmUnsupervisedGraphAI}, an idea that we draw upon.
However, unlike our work, prior systems detect lateral movement by applying narrow signatures or traditional machine learning techniques to flag anomalous activity.
Kent et al.~\cite{kent2015authentication} detect the use of compromised credentials by training a logistic regression model to detect when an account accesses an unusual set of machines;
their classifier achieves a true positive rate of 28\% and incorrectly flags 1 / 800 users as compromised.
Bowman et al.~\cite{lmUnsupervisedGraphAI} and log2vec~\cite{liu_log2vec_2019} use deep-learning methods to build anomaly detection systems, with hand-tuned thresholds, that identify clusters of suspicious logins.
These approaches incur false positive rates ranging from 0.9\%~\cite{lmUnsupervisedGraphAI} to 10\%~\cite{liu_log2vec_2019} to detect 80--90\% of simulated attacks and/or red team exercises in their data.

Among the best performing prior work, Siadati and Memon propose a detector for identifying ``structurally anomalous logins'', which we refer to as SAL~\cite{ccs2017lmdetecting}.
On one month of data, SAL can detect 82\% of randomly generated attack logins at a 0.3\% false positive rate ($>$~500 false alarms/day on their dataset).
Whereas SAL focuses on identifying point-wise anomalous logins (``one-hop'' paths),
Latte~\cite{liu2018latte} detects two-hop lateral movement attacks
by identifying paths where each login has rarely occurred in prior history.
Latte then uses a specific signature to reduce false positives by only alerting on
rare paths that also include a remote file execution operation on the path's final machine (identified by a set of hard-coded Windows events).
Based on one day of data and a specific anomaly threshold, Latte can detect a pentester exercise while generating 13 false alarms.
Although Latte can identify longer attack paths, its narrow signature, which requires the attacker to perform a specific action on the final host, can lead to false negatives.
Moreover, implementing this signature faces practical challenges,
since common authentication logs from Linux and Mac OS systems do not provide an easy way to re-implement Latte's Windows-specific signature.

Although they provide good starting points for detection, prior systems generate an impractical volume of false positives or incur too many false negatives (Section~\ref{sec:evaluation:baseline} reports the performance of SAL on our data set).
Our work addresses these challenges with a new approach to identifying suspicious login paths. 
Rather than alerting on paths that are simply anomalous or relying on signatures that target specific host operations,
we identify a set of key properties about attack paths based on the overarching goals of lateral movement.
By focusing on paths with these properties, and only applying anomaly detection in scenarios with high uncertainty,
our approach detects a wider range of attacks than those that employ a narrow signature, while also generating fewer false positives than traditional anomaly detection methods.

\subsection{Security Model}
\label{sec:threatmodel}

\paragraph{Detection Goals}
\detector aims to (1) detect a diverse range of lateral movement attacks, while (2) generating a very low volume of false positives.
We focus on developing detection for settings where an organization has a team of security analysts with a limited time budget for reviewing alerts.
In particular, we design \detector to score a set of movement paths in terms
of how problematic the activity appears to be,
allowing an organization to specify their own bound on the number of alerts that \detector generates. 
Based on prior work~\cite{ho2017detecting,usenix_ti} and the practical experiences of security team members at \dropbox, this alert-budget design accurately reflects a real-world operating model for many organizations.
We consider \detector successful if it produces an alert for any login made by an attacker. 
Upon confirming the presence of an attack, organizations can use forensic techniques from complementary work~\cite{hossain2018dependence,hassan2020omegalog,wilkens_towards_2019} to perform further analysis and remediation.

\paragraph{Threat Model}
Similar to prior work, we focus on detecting interactive and credential-based lateral movement attacks~\cite{ccs2017lmdetecting}.
Under this threat model, we assume that an attacker has managed to compromise an initial ``foothold'' machine within the enterprise,
but they (1)~need to acquire additional credentials to access the data or systems they ultimately seek,
and
(2)~move between machines via login or remote command execution events that use a set of credentials for authentication.
In particular, attackers may exploit vulnerabilities on machines or weak authentication protocols (\eg privilege escalation or pass-the-hash attacks), but we assume that their movement between machines produces a login event visible to our detector.
Additionally, this threat model focuses on attackers who manually perform the movement (login) operations during their attack,
as opposed to an attack that installs malware that moves to new systems autonomously.
Our threat model reflects the behavior of many real-world lateral movement attacks,
ranging from targeted attacks by state-sponsored actors~\cite{bronzeunion, operationsmn,fireeyeApt1,lateralMovementNCSC,lateralMovementUSCert,sshLateralMovementIran,sshLateralMovementChina} to newer and stealthier forms of ransomware~\cite{humanOperatedLM1, humanOperatedLM2}.

\begin{table}[t]
  \small
	\begin{center}
    	\begin{tabular}{ll}
    	\toprule
    	\textbf{Nodes (Source + Destination Machines)} & \textbf{Edge (Login)}\\
    	\midrule
        Hostname & Timestamp \\
        Client vs. server &  Target username \\
    	Owner's username (clients only)  & \\
      \bottomrule
    	\end{tabular}
\setlength{\belowcaptionskip}{-18pt}
	\caption{The information for each login event in our data.
	Each login creates a unique edge between two nodes (internal machines) in the graph that \detector constructs
	(\S~\ref{sec:overview:architecture}).
	}
    \label{table:data:graph_attributes}
	\end{center}
\end{table}

\section{Data}
\label{sec:dataset}

Our work uses a collection of successful login events between internal machines by employees at \dropbox\footnote{Because our work focuses on mitigating successful lateral movement, our analysis omits failed logins; however, future work could investigate ways to incorporate such failures as additional detection signals.},
a large enterprise that provides storage and cloud collaboration services to hundreds of millions of users.
Whenever a machine receives a remote access attempt from another machine (\eg an inbound ssh session or a remote command execution issued via utilities like psexec),
the receiving machine generates a record of a remote ``login''.
Because most operating systems record these login events by default, organizations collect these authentication logs as part of standard security best practices.
This data provides visibility into the internal logins between machines within \dropbox's corporate network,
such as client laptops, authentication servers (\eg Windows Domain Controller), and a variety of infrastructure and application servers (\eg DNS servers, machines that test and build applications, and analytics servers).
Representative of the heterogeneous nature of modern enterprises, the logins in our data span a variety of authentication protocols (\eg Kerberos and ssh) across many types of devices (laptops, physical servers, and virtual machines), operating systems (Windows, Mac OS, and Linux),
and account types (\eg regular users, administrators, and service accounts).

\subsection{Data Size and Schema}
\label{sec:data:login_filtering}
Our data contains \numRawLogins successful logins from \dateStart to \dateEnd (\totalMonths months).
As shown in Table~\ref{table:data:graph_attributes}, each login event contains a timestamp, the target username of the login,
the source and destination machines that initiate and receive the login, respectively, and metadata about these machines.
These logins span \numUsers accounts and occur between \numMachines machines.
Section~\ref{sec:generalizability} provides more details about the graph topology of our login data,
and how different network configurations might affect our detection algorithms.

\subsection{Data Cleaning}
\label{sec:data:cleaning}
The vast majority of our data's login events do not reflect meaningful remote access events (\ie did not enable a user to remotely execute commands or access sensitive data on the destination machine).
\detector applies four filtering rules described below to remove these logins from our data set.
Excluding these spurious logins, our data set contains \numFinalLogins successful logins, with a median of \medianDailyLogins logins per day.

\paragraph{Filtering Windows logins}
As noted in prior work~\cite{kent2015authentication}, many ``logins'' between internal machines in Windows enterprise environments do not represent a meaningful remote access event.
Rather, these logins often correspond to uninteresting artifacts and special API calls that result from Windows enterprise logging, and do not provide a user with the ability to access data or alter the destination machine.
Removing these logins from our data results in a \reductionFactorWindows reduction,
which comes primarily from removing three types of logins: printing jobs, authentications into update and logging servers, and non-administrator logins to Windows Domain Controllers.
Most non-administrator logins to Domain Controllers correspond to artifacts of Kerberos authentication,
where Domain Controllers serve the role of a Kerberos Key Distribution Center (KDC)
and requests for a Kerberos ticket generate a record of a ``login'' into the Domain Controller.
After removing this collection of spurious logins,
our data set contains roughly \numTotalLoginsAfterWindows login events.

\paragraph{Filtering automation logins}
We further winnow our data set by removing internal logins that result from low-risk automation.
\detector analyzes a historical set of logins and identifies a set of login edges that correspond to automation.
Specifically, each automation edge consists of a triplet (source, destination, and username),
that (1) occurs frequently across our data,\footnote{In our work, we define a frequently occurring edge as one that occurs greater than $N = 24\times D$ times, where $D$ equals the number of days in the historical data set (i.e., in total, the edge occurs at least as often as a process that runs once every hour on each day in the historical data set).}
(2) occurs on at least 50\% of the historical days, and
(3) has a target username that does \textit{not} match any employee's account (\ie a non-human username).
\detector then outputs a list of these edges as candidates for automation related logins.
After a review by the organization's security team, \detector removes any login whose (source, destination, and target user) matches an edge listed in the approved automation set.

In our data, \detector identifies a set of approximately \numAutomationEdges automation edges that account for over \numAutomationLoginsApprox login events.
Manually inspecting these automation logins reveals that they correspond to mundane operations with minimally privileged service accounts via a restricted set of remote-API calls (\eg specific remctl calls~\cite{remctl} exposed by the destination machines).
For example, many of these logins resulted from file synchronization operations between a central ``leader'' node and geographic replicas (\eg a central software repository machine syncing its content with replicated, regional servers).
Another common category of these automation logins corresponds to version control and bug tracking software performing git operations to synchronize state among each other;
these internal logins occurred under a restricted ``git'' user account that has access to a limited API of git operations.

\subsection{Ethics}
\label{sec:ethics}
This work involved a collaboration between academia and industry.
Our research used an existing, historical data set of employee logins between internal machines at \dropbox, which enterprises commonly collect to secure their environment.
Only authorized security employees at \dropbox accessed this data;
no sensitive data or personally identifying information was shared outside of \dropbox.
Additionally, the machines that store and operate directly on data from \dropbox's customers reside on separate infrastructure;
our study did not involve that infrastructure or access any customer-related data.
This project underwent internal review and received approval by the legal, privacy, and security teams at \dropbox.

\section{Modeling Lateral Movement}
\label{sec:overview}

\paragraph{Our Approach}
\detector, our system, constructs a graph of user logins between internal machines and then detects lateral movement by identifying suspicious paths in this graph.
A suspicious path corresponds to a sequence of logins made by a single actor with two properties:
(1) the path has at least one login where the actor uses a set of credentials that does not match their own,
(2) the path accesses at least one machine that the actor does not have access to under their own credentials.

\paragraph{Motivating Intuition}
This approach leverages a simple yet powerful observation:
in many real-world enterprise attacks, adversaries conduct lateral movement to acquire additional credentials and access new machines that their initial foothold did not have access to~\cite{bronzeunion, operationsmn, fireeyeApt1,cobaltKitty,lateralMovementNCSC,sshLateralMovementIran,sshLateralMovementChina}.
For example, at many organizations, access to sensitive data and/or powerful internal capabilities requires a special set of privileges, which most enterprise users lack.
Thus, attacker lateral movement will produce paths that use a new (elevated) set of credentials (Property 1) and access sensitive machines that their initial victim could not access (Property 2).
By searching for these two key properties, \detector also illustrates how login data not only provides visibility into attacker lateral movement,
but also contains latent signals that reveal the completion of other core stages of an attack's lifecycle.
For example, Property 1 captures the fact that attackers frequently acquire privileged credentials (the ``privilege escalation'' and ``credential access'' stages from the MITRE ATT\&CK Framework~\cite{mitreattack}) to access additional machines within an organization.

Moreover, the combination of these two attack path properties corresponds to characteristics that we do not expect in benign paths:
users should access machines under their own credentials and they should only login to machines that they have legitimate privileges to access.

\subsection{Challenge: Anomalies at Scale}
\label{sec:overview:naive_anomaly}

Prior work detects lateral movement by identifying logins that traverse rare graph edges,
under the assumption that attacker movement will occur between users and machines that rarely interact with each other~\cite{liu2018latte,ccs2017lmdetecting,bohara2017unsupervised}.
While intuitive, these approaches generate too many false positives,
due to the volume of rare-but-benign behavior that occurs in large enterprises.

Even after applying \detector's data cleaning steps (\S~\ref{sec:data:login_filtering}),
tens of thousands of logins create ``rare'' graph edges in our data set.
If we alerted on logins whose edges have never occurred in recent history,
such a detector would produce over \numNewEdgeAlertsApprox alerts across our data (over 1,600 alerts / month).
These rare-but-benign logins stem from a diverse set of causes, such as
users performing maintenance on machines they rarely access (\eg a user serving on their team's on-call rotation), new users or employees returning from a long vacation, and users simply accessing rare-for-their-role services.
Although prior work introduces techniques to refine this anomaly detection approach,
they still produce too many false positives (\S~\ref{sec:evaluation:baseline}).
By re-framing the definition of an attack path from simply anomalous paths, to paths that contain the key properties we highlight, \detector can detect a range of lateral movement attacks with significantly fewer false positives.

\begin{figure}[t]
\centering
\includegraphics[width=0.98\columnwidth]{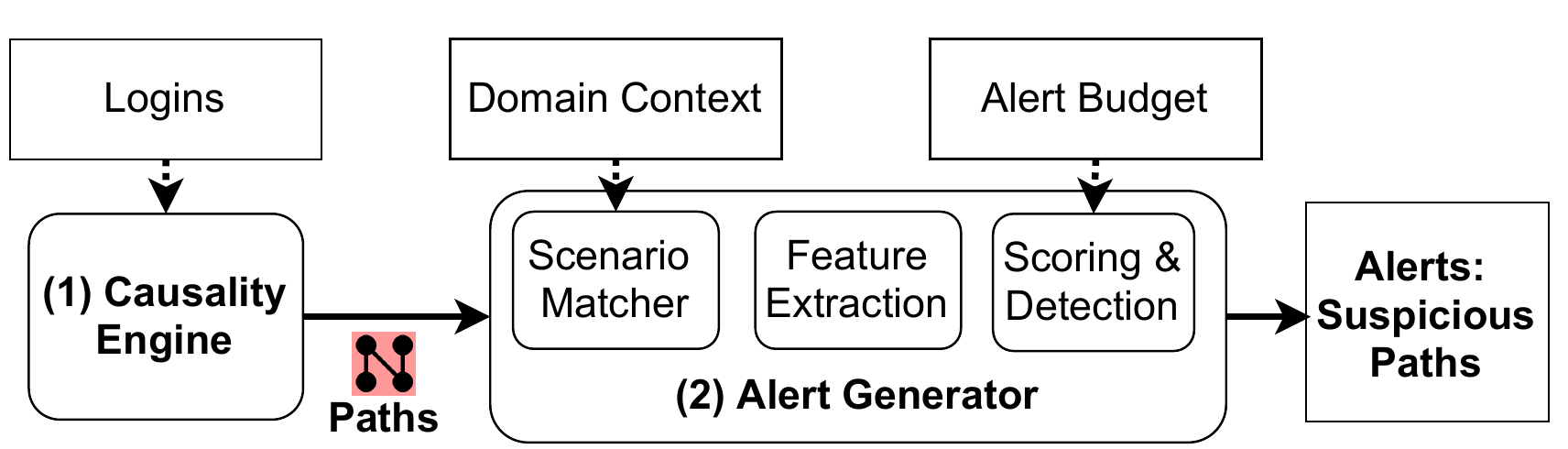}
\setlength{\belowcaptionskip}{-10pt}
\caption{\detector analyzes login events between internal machines within an enterprise and generates alerts
for paths of logins that correspond to suspicious lateral movement activity.
\detector has two key components:
(1) a causality engine that infers a set of causal paths that a login might belong to (\S~\ref{sec:paths}), and
(2) detection and scoring algorithms that decide whether to alert on a path of logins (\S~\ref{sec:detector}).
}
\label{fig:overview}
\end{figure}

\begin{figure}[t]
\centering
\includegraphics[width=0.99\columnwidth]{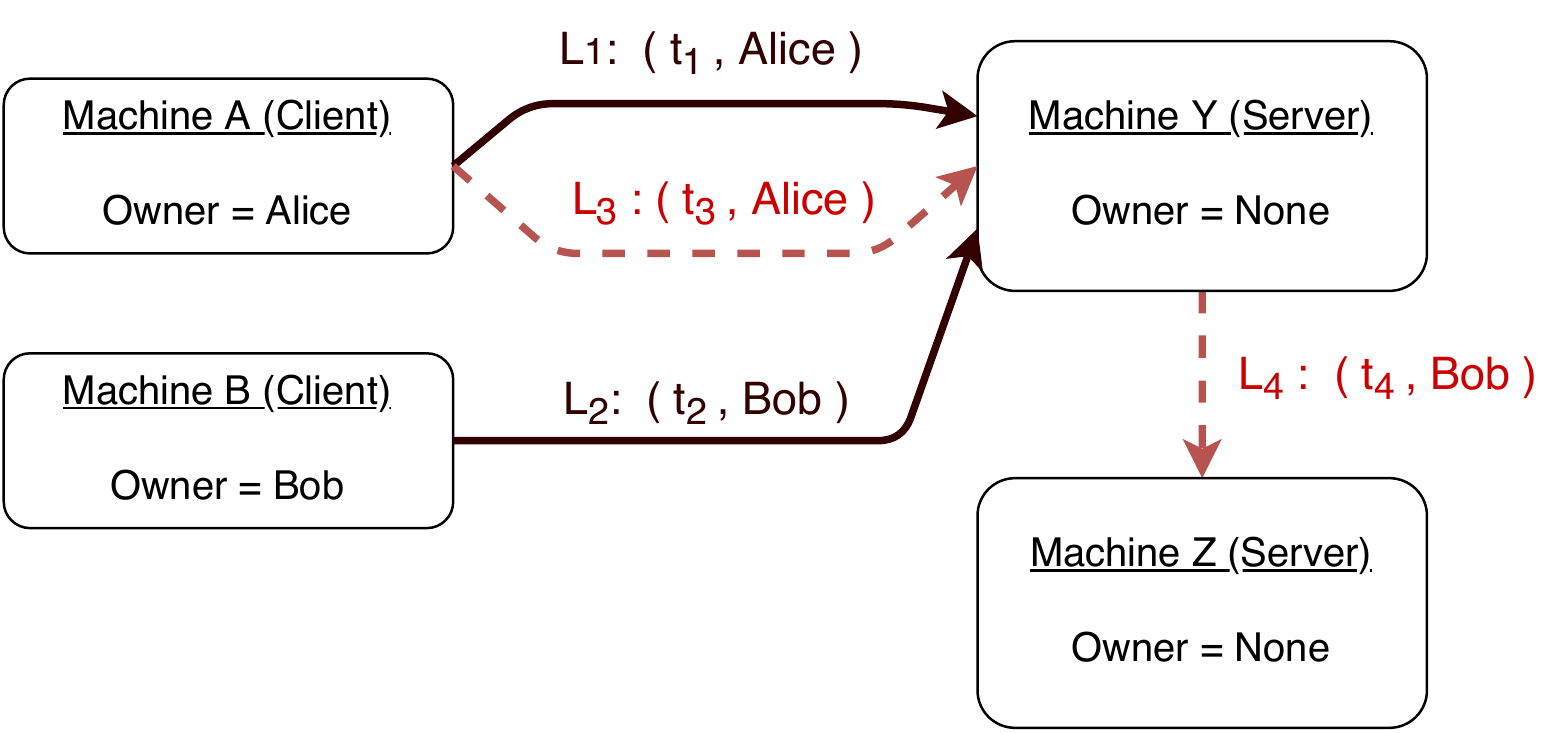}
\setlength{\belowcaptionskip}{-10pt}
\caption{An example of a simple login graph.
Solid black edges ($L_1$ and $L_2$) correspond to benign login events.
Dashed red edges ($L_3$ and $L_4$) correspond to a lateral movement attack path.
}
\label{fig:path_example}
\end{figure}

\subsection{\detector: System Overview}
\label{sec:overview:architecture}
\detector consists of two stages, shown in Figure~\ref{fig:overview}.
The first stage of \detector (\S~\ref{sec:paths}) runs a ``causality engine'' that aggregates a set of logins into a graph of user movement
and identifies broader paths of movement formed by groups of logically-related logins.
The second stage of \detector (\S~\ref{sec:detector}) takes a set of login paths and decides whether to generate an alert by identifying which login paths contain the two key attack properties described above.
During this final stage, \detector prunes common benign movement paths, extracts a set of features for each path, and uses a combination of detection rules and a new anomaly scoring algorithm to compute the ``suspiciousness'' of each login path.

\paragraph{The Login Graph}
Given a set of logins, \detector constructs a directed multi-graph that captures the interactions among users and internal machines. 
Figure~\ref{fig:path_example} shows a simple example of a login graph constructed by \detector.
Each login creates a directed edge in the graph,
where the edge's source and destination nodes correspond to the machine initiating and receiving the login. 
Edges represent unique, timestamped logins from the source to the destination machine;
multiple logins between the same two machines generate multiple edges.
Each edge is annotated with a target username: the account that was logged into on the destination machine
(the username and permissions that the new session operates under).

\paragraph{Login Paths and Causal Users}
A path of logins corresponds to a series of connected edges, where each edge is ``caused'' by the same actor.
We use the term \textit{causal user} to refer to the actor whose machine initiated a path of logins,
which might not be the same as the \textit{target user} recorded in each login.
The causal user is the original actor responsible for making these logins (taken from the first edge in each path), while each login's target user reflects the credentials that the login's destination machine received.

For example, in Figure~\ref{fig:path_example}, an attacker compromises Alice's machine (\pathStart) and makes a series of internal logins that forms a two-hop lateral movement path from Machine \pathStart to \targetMachine.
The attacker first uses Alice's credentials in a login to Machine $Y$, shown as $L_3$.
Then the attacker compromises Bob's credentials on \switchingSrc
and uses them to login to Bob's account on $Z$, labeled $L_4$.
For each of the logins in this path, Alice is the causal user,
since all of the logins were made (caused) by a user starting from Alice's machine.
Alice and Bob are the target users of $L_3$ and $L_4$ respectively,
since each login presented those usernames and credentials during authentication.

\paragraph{Path Types}
One of the key attack properties that \detector looks for is whether a path's causal user ever authenticates
into a machine with a new set of credentials.
As described later in Section~\ref{sec:paths}, the information provided in standard authentication logs does not always enable \detector to precisely infer whether a path exhibits this property.
Accordingly, \detector makes a distinction between three types of paths:
a \textsc{Benign} path, a path with a \textsc{Clear} credential switch, or an \textsc{Unclear} path.

\detector labels a path as \textsc{Benign} if every login in the path uses the causal user's credentials (\eg no switch in credentials occurred).
A path has a \textsc{Clear} credential switch if at least one login in the path must have switched to a new set of credentials.
For example, in Figure~\ref{fig:path_example}, assume that login $L_2$ did not occur at all,
then the paths ($L_1$, $L_4$) and ($L_3$, $L_4$) correspond to paths with a \textsc{Clear} switch,
because all paths leading to $L_4$ previously used a different set of credentials.
On the other hand, if all of $L_1$, $L_2$, $L_3$ occurred and \detector cannot clearly determine which of them caused $L_4$,
then \detector will treat both the paths ($L_1$, $L_4$) and ($L_3$, $L_4$) as \textsc{Unclear} paths.
An \textsc{Unclear} path corresponds to a situation where \detector cannot cleanly infer a causal path for a given login,
but rather infers multiple potential paths, where some of the paths involve a switch in credentials (\eg $L_3$ to $L_4$),
but others do not (\eg $L_2$ to $L_4$).
As discussed in Section~\ref{sec:detector}, because of these different levels of certainty,
\detector uses two sets of detection algorithms to classify a path as malicious.
For paths with a \textsc{Clear} credential switch, \detector applies a simple rule-set (\S~\ref{sec:detection:detector_switch_clear}).
However, when limitations in real-world logs create uncertainty about the paths that \detector's causality engine infers (\ie \textsc{Unclear} paths),
\detector uses an anomaly scoring algorithm to determine when to alert on a path (\S~\ref{sec:detection:detector_switch_unclear}).

\section{Inferring Causal Login Paths}
\label{sec:paths}

Standard authentication logs describe point-wise activity that lacks broader context about each login,
such as from whom and where the login originated.
For example, in Figure~\ref{fig:path_example}, given login $L_4$ in isolation,
a detector does not know whether $Bob$ accurately reflects the user responsible for making the login, or whether another user such as \causalUser has stolen Bob's credentials and used them in a malicious login.
Thus, for each login (\newLogin) that occurs,
the first stage of \detector runs a ``causality engine'' that coarsely infers the broader path of movement that a login belongs to and the \textit{causal user} responsible for initiating the movement path.
To do so, \detector uses a time-based heuristic to infer a set of ``causal paths'' for \newLogin,
where each path corresponds to a unique sequence of connected logins that could have led to \newLogin and occurred within the maximum time limit for a remote login session.

\begin{table}[t]
  \small
	\begin{center}
    \resizebox{\columnwidth}{!}{
    	\begin{tabular}{ll}
    	\toprule
    	\textbf{Path Component} & \textbf{Description} \\
    	\midrule
    	Login List & List of logins in the path \\[0.04in] 
    	Causal User & Username of the employee whose \\
    	& machine initiated the path \\[0.04in]
    	Changepoint Logins & A list of logins where the username \\
    	&  differs from the path's preceding login \\[0.04in]
    	Path Type & \textsc{Benign}, \textsc{Clear}, or \textsc{Unclear}: whether \\
    	& the path switches to new credentials \\[0.04in]
      \bottomrule
    	\end{tabular}
    }
\setlength{\belowcaptionskip}{-20pt}
	\caption{Information in each path generated by \detector's causality engine (\S~\ref{sec:paths}).
	Given a new login, \detector infers a set of these causal paths, each of which reflects a sequence of logins that an actor could have made up to and including the new login.
}
    \label{table:path:schema}
	\end{center}
\end{table}

\paragraph{Identifying Causally-Related Logins}
\detector produces a set of causal paths by running a backwards-tracing search from \newLogin to identify a sequence of causally-related logins that include \newLogin.
Two logins are causally related if they (1) form a connected set of edges in the login graph and (2) occur within $T$ hours of each other.
Concretely, we say that \otherLogin is a causal, inbound login for \newLogin if the destination of \otherLogin equals the source machine of \newLogin, and \otherLogin occurred within 24 hours prior to the time of \newLogin.
We choose a threshold of 24 hours based on the maximum duration of a login session at \dropbox;
for sessions that exceed this duration, the company requires the source machine to re-authenticate, which produces a fresh login event in our data.
For example, in Figure~\ref{fig:path_example}, $L_1$, $L_2$, and $L_3$ are all causal logins for $L_4$ if they occurred within 24 hours prior to $t_4$.
Using this causal rule, \detector infers a set of login paths by identifying all of the causal logins for \newLogin,
and then recursively repeats this search on each of those causal logins.

This process is similar to provenance and taint-tracking methods that trace the flow of information from a sink (\newLogin's destination machine) back to its source (the root node of \newLogin's login path)~\cite{hassan2020tactical,hossain2020combating,hossain2018dependence}.
As with these flow-tracking methods, naive backwards-tracing risks a ``dependency explosion'',
where each backwards step can exponentially increase the number of paths that \detector infers,
but only one of these paths represents \newLogin's true causal path.
We find that four optimizations and environmental factors mitigate this risk.

First, \detector can use an optimized implementation that requires only a single-step of backwards-tracing per login.
At a high-level, based on our key attack properties,
\detector only needs to analyze paths that involve a switch in credentials (Property 1).
As a result, \detector can incrementally build a set of ``watchlist'' paths that contain a potential switch in credentials.
For each new login, \detector only needs to perform one step of backwards-tracing to determine if the new login involves a switch in credentials, or if it extends one of these watchlist paths;
Appendix~\ref{appendix:implementation:paths} describes this implementation in more detail.
Second, we observe that enterprise networks tend to have a relatively flat topology,
since most users prefer to directly access their target server;
this behavior limits dependency explosion, which we discuss more in Section~\ref{sec:generalizability}.
Third, due to the natural workflows of users and a standard implementation of least privileges, most machines only get accessed by a handful of users for specific job duties.
This clustering limits the number of inbound logins per machine, which reduces the potential for path explosion (\S~\ref{sec:generalizability}).
Finally, to mitigate path explosion that can occur from users or scripts making many repeated logins to/from a machine,
\detector deduplicates paths to one unique path per day
(\ie one unique set of daily login edges, where a daily edge is a four-tuple of a login's source, destination, target username, and timestamp rounded to the date it occurred).

\paragraph{Path Components and Types}
Every causal path inferred by \detector contains the information in Table~\ref{table:path:schema}.
Each path includes a list of ``changepoint'' logins:
logins that used a different username than the preceding login in the path.
For logins that occurred from a client source machine, if the target username does not match the source machine's owner, \detector also adds this login to its changepoint list.

\detector computes a path's \textit{causal user} by examining the first (earliest) login in the path.
If the login's source machine is a server, then \detector treats the target username as the path's causal user.
However, if the first login's source machine is a client, \detector takes the owner of that source machine and treats that username as the causal user:
clients typically correspond to the start of a user's movement path and logins from these machines should use their owner's credentials.
Additionally, \detector takes a user-provided list of special ``bastion'' machines:
hardened gateway servers that provide access to restricted network segments or machines,
and which require users to perform heightened authentication to access these protected parts of the network
(\eg password and hardware-based 2FA authentication during each login).
Whenever \detector encounters a login that originates from a bastion source machine,
it treats this login as the root login for the path: \ie~\detector treats the username of the bastion login as the path's causal user,
and stops performing backwards-tracing for the path.
Because bastions require robust forms of authentication,
logins forwarded from bastion source machines (\ie logins that successfully authenticated to the bastion server)
indicate that the login's purported username does reflect the true actor responsible for making the login.

Paths belong to one of three types: a \textsc{Benign} path, a path with a \textsc{Clear} credential switch, or a path with \textsc{Unclear} causality.
For each changepoint login in a path,
\detector checks whether the changepoint login's username matches any of the usernames across its potential inbound (causal) logins.
If all of the inbound hops used a different username, or if the changepoint login originated from a client source machine,
then the path has a \textsc{Clear} credential switch; otherwise, \detector labels the path as \textsc{Unclear}.
If a path does not have any changepoint logins, then \detector marks the path as \textsc{Benign}.

For example, in Figure~\ref{fig:path_example}, if $L_1$, $L_2$, and $L_3$ occurred within 24 hours prior to $L_4$, \detector will produce 3 causal paths for $L_4$.
The paths starting with $L_1$ and $L_3$ will form \textsc{Unclear} paths, and the path starting with $L_2$ will get marked as \textsc{Benign}.
The path from $L_2$ to $L_4$ will list $Bob$ as its causal user and have no changepoints logins.
Both the attack path ($L_3$ to $L_4$) and the path from $L_1$ to $L_4$ will list \causalUser as their causal user,
and contain $L_4$ in their list of changepoint logins.

\begin{figure*}[t]
\centering
\includegraphics[width=0.9\textwidth]{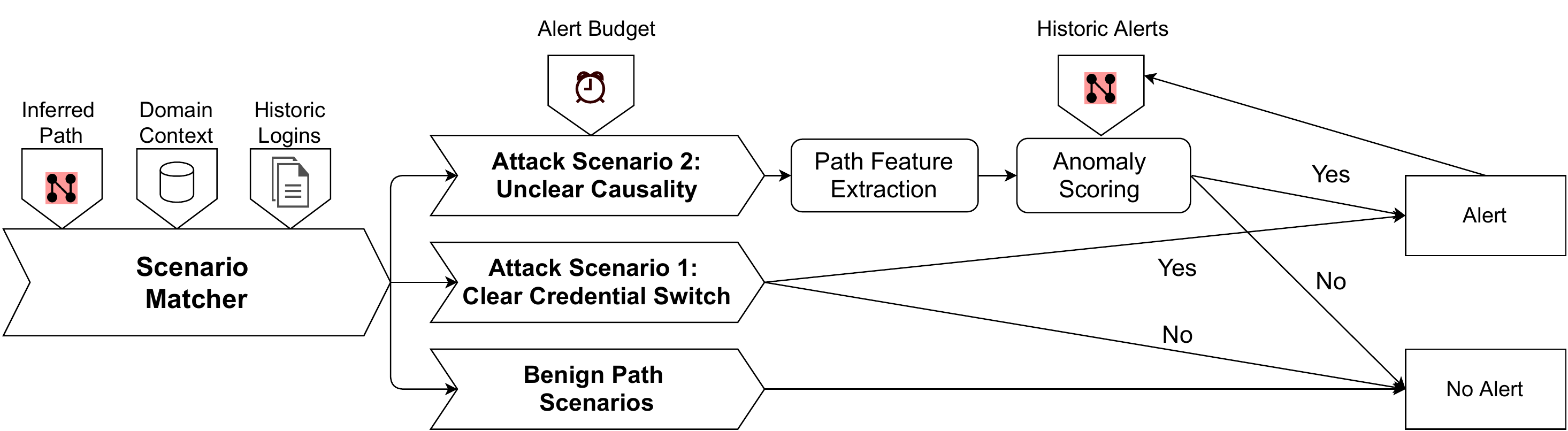}
\caption{Architecture of \detector's alert generator (\S~\ref{sec:detector}).
Given a login path (\S~\ref{sec:paths}),
\detector checks whether the path matches a benign scenario or an attack scenario.
Based on the path's scenario, \detector either discards the path or generates an alert if the scenario's detector triggers.
}
\label{fig:detection}
\end{figure*}

\section{Detection and Alerting}
\label{sec:detector}

\detector classifies each path given two additional inputs:
a set of historical logins for feature extraction
and a user-provided ``budget'' that controls the daily number of alerts that \detector produces for \textsc{Unclear} paths (\S~\ref{sec:detection:detector_switch_unclear}).
\detector first checks whether the path matches one of five benign scenarios;
if so, it does not generate an alert.
For paths that do not match a benign scenario,
\detector identifies which of two attack scenarios the path might belong to and applies the scenario's corresponding detector.
These detectors apply either a rule set (\S~\ref{sec:detection:detector_switch_clear}) or an anomaly scoring algorithm (\S~\ref{sec:detection:detector_switch_unclear}),
and produce an alert if the path is marked as suspicious.

\paragraph{Benign Movement Scenarios}
In the first benign scenario, \detector marks a path as benign if every one of its logins uses its causal user's credential (\ie a path labeled as \textsc{Benign} by the causality engine);
because these paths do exhibit the first key attack property, \detector discards them.
\detector also labels approximately \numTotalBenignPruningLogins paths as benign if they match one of four other benign and low-risk scenarios.

First \detector identifies one-hop paths (\ie logins) from new machines and new users:
\detector labels the path as benign if either the user and/or source machine have existed for less than one week
(based on their earliest occurrence in historical logins and the organization's inventory databases).
Second, \detector ignores all paths that originate from a machine undergoing provisioning for a new owner.
As part of this process, an administrator runs a script that authenticates into several specialized servers to configure the machine
(\eg installing the operating system and configuring the new owner's account).
These logins will seem suspicious to \detector because they will use an administrator's credentials (target username) that differs from the machine's owner (the causal user).
To identify login events that relate to machine re-provisioning, \detector checks for three properties:
(1) the login's destination belongs to a set of dedicated provisioning servers,
(2) the login's target user is a system administrator, and
(3) the login originates from a dedicated subnet used for machine provisioning.
If \detector encounters a login with these three properties, it does not run its causality engine or generate an alert.
In total, \detector removes approximately 125,000 logins related to new machines or those undergoing provisioning.

Third, the use of (non-human) service accounts produces roughly 42,000 one-hop paths that \detector would otherwise label as cases of clear-credential switching.
In these logins, a legitimate user performed a login using a ``mismatched'' set of credentials that correspond to a service account;
however, these logins correspond to an expected credential switch, as part of accessing a service within the enterprise.
For example, these logins includes users running a script to launch testing jobs when building a new version of \dropbox's desktop application;
part of this script includes remote commands issued to the build and test machines under a service account (\eg user = \textit{test-services}).
\detector infers a set of these service usernames by identifying any username that (1) does not match an employee username,
and (2) was used in successful logins from more than ten different source machines across a set of historical data.
To ensure that usernames inferred by \detector do not provide widespread access or highly privileged capabilities,
\detector outputs the set of inferred service accounts for an organization's security team to confirm, and uses only the set of approved service usernames when filtering these benign logins.
Because these accounts are designed for a limited and specific service operation,
organizations can mitigate the risk of lateral movement via these credentials by configuring them with a limited set of permissions to a specific set of machines;
at \dropbox, many of these service accounts also access their destinations via a limited remote command API~\cite{remctl}, as opposed to creating a full interactive session.

The final benign scenario involves logins to and from a bastion host.
Organizations often segment parts of their network for improved efficiency, maintenance, and security by placing a set of machines behind a hardened bastion host~\cite{wikipediaNetworkSegmentation, ciscoNetworkSegmentation}.
To access a server within this network segment, a user must first tunnel and authenticate through the network segment's bastion.
\dropbox's corporate network contains a few such network segments.
Because bastion machines correspond to hardened hosts, perform a limited set of operations (authentication and connection forwarding), and often do not allow users to establish logins onto the host itself,
a login that originates from a bastion likely reflects legitimate user activity.
Given a list of bastion hosts at an organization, \detector does not alert on any one-hop path that originates from a bastion or any two-hop paths that traverse a bastion.

\paragraph{Attack Scenarios}
If a path does not match any of these benign scenarios, \detector checks whether it matches one of two attack scenarios and, if so, applies the corresponding detection algorithm to see whether it should produce an alert.
First, if the path contains a login that switches credentials and the causality engine has high confidence that the switch occurred (a \textsc{Clear} path),
\detector applies a simple rule set to classify the path as suspicious or not (\S~\ref{sec:detection:detector_switch_clear}).
However, because of imperfect information contained in real-world authentication logs,
\detector's causality engine sometimes infers multiple potential paths that a login could belong to, where not all of the paths contain a credential switch (\ie paths with \textsc{Unclear} causality).
Because of this uncertainty,
\detector's second detector evaluates how suspicious each such path is with a probabilistic scoring algorithm (\S~\ref{sec:detection:detector_switch_unclear}) and alerts if the path has one of the most suspicious scores in recent history.

\subsection{Attack Scenario \numScenarioClearSwitch: Paths with a Clear Credential Switch}
\label{sec:detection:detector_switch_clear}

Paths with a clear credential switch contain at least one login where \detector knows that the causal user it inferred for the path must have switched to a different set of credentials (the first key attack property).
For these paths, \detector generates an alert if the path accesses any destination that its causal user has never accessed in prior history;
a conservative estimate of when a path's causal user accesses an unauthorized machine.

More formally, let \pathCurrent represent a path with a causal user of \causalUser and
\pathDestList refer to the destination machines across all of \pathCurrent's logins.
\detector generates an alert if \pathCurrent exhibits the two key attack properties:
\begin{enumerate}
    \item Property 1: \pathCurrent has a \textsc{clear} credential switch (path type).
    \item Property 2: \pathCurrent contains at least one destination in \pathDestList that \causalUser has never accessed in the historical training data (\eg past 30 days).
\end{enumerate}

\subsection{Attack Scenario \numScenarioUnclearSwitch: Paths with Unclear Causality}
\label{sec:detection:detector_switch_unclear}
The second attack scenario handles paths with \textsc{Unclear} causality:
when \detector infers multiple causal paths for a login, where some paths contain a credential switch and others do not (\S~\ref{sec:paths}).
To handle unclear paths, \detector uses a probabilistic detection algorithm to identify and alert on paths that are highly anomalous.
This selective use of anomaly detection, only in cases where the limitations of authentication logs introduce uncertainty about whether a path contains the key attack properties, distinguishes \detector from prior work, which
simply applies anomaly detection to every path.

\paragraph{Alert Overview: Unclear Causality}
Given an \textsc{Unclear} path (\pathCurrent), \detector first checks whether the path ever visits a machine that its causal user (\causalUser) has not previously accessed in the training data (the second attack property).
If \causalUser has access to all of the path's destinations, then \detector marks the path as benign.\footnote{Future logins in the path will cause \detector to produce extended paths that its detection algorithm will subsequently examine.}
Otherwise, \detector runs the following anomaly detection algorithm on \pathCurrent.

First, \detector extracts three features that characterize \pathCurrent's rareness.
Next, \detector uses \pathCurrent's features to compute a ``suspiciousness'' score for the path,
which it then uses to rank \pathCurrent relative to a historical batch of paths (\eg the past 30 days).
If \pathCurrent ranks among the top $30\times$\budgetSym most suspicious historical paths, then \detector generates an alert.
\budgetSym corresponds to a user-provided budget that specifies the average number of daily alerts that an analyst has time to investigate for these types of attack paths.

\paragraph{Path Features}
\detector uses a set of historical ``training'' logins to extract three features for a path.
Let \pathStart refer to the path's starting machine and \pathEnd refer to the path's final destination.
Given a path's changepoint login (\changepointLogin),
\detector computes two numerical features.
First, \detector computes the historical edge frequency for each login preceding \changepointLogin,
where an edge's historical frequency equals the number of \textit{days} that a successful login with the exact same edge (source, destination, and target username) has occurred in the training data;
the first feature value equals the minimum (lowest) frequency among these preceding logins.
Second, \detector computes the historical edge frequency for each login in the remainder of the path,
and takes the lowest frequency value among these hops;
\ie the historical frequency of the rarest login starting at \changepointLogin until the path's final hop.
For the third feature, \detector computes the number of historical days where any successful login path connects Machine \pathStart and Machine \pathEnd. 
If a path has multiple changepoint logins, \detector computes these three features for each changepoint login,
runs its anomaly scoring algorithm (below) for each feature set,
and then uses the most suspicious score for the path.

\newcommand{\weightedSum}{Sum$_F$\xspace}
\newcommand{\subscore}{Sub-Score$_F$\xspace}
\newcommand{\pathscore}{PathScore\xspace}
\begin{algorithm}[t] 
    \begin{algorithmic}
    \item[] AlertGen($P$, $A$ (historical alerts), $L$ (historical paths)):
    \item[1:] \textbf{for} each path $X$ in $A$  \textbf{do:}
    \item[2: \hspace{3mm}] \textbf{if} Score(\pathCurrent, $L$) $\geq$ Score($X$, $L$):
    \item[3: \hspace{6mm}] Alert on \pathCurrent
    \newline
    \item[] Score(\pathCurrent, $L$): $\displaystyle\prod_{F}$ Sub-Score(\pathCurrent, $L$, $F$)
    \newline
    \item[] Sub-Score(\pathCurrent, $L$, $F$ (feature)):
    \item[1:] \weightedSum $\gets$ 0
    \item[2:] $N$ $\gets$ 0 (the total \# of true causal paths)
    \item[3:] \textbf{for} each path $X$ in $L$ \textbf{do:}
    \item[4: \hspace{3mm}] \textbf{if} \pathCurrent has a smaller value for $F$ than $X$:
    \item[5: \hspace{6mm}] \weightedSum $\gets$ \weightedSum + $C_x$
    \item[\hspace{10mm}] where $C_x$ = the path certainty for $X$ (\S\ref{sec:detection:detector_switch_unclear})
    \item[6: \hspace{3mm}] $N$ $\gets$ $N$ + $C_x$,
    \item[7:] \subscore $\gets$ \weightedSum~/ $N$
    \end{algorithmic}
\caption{\detector's anomaly scoring algorithm}
\label{alg:scoring}
\end{algorithm}

\paragraph{Anomaly Scoring}
Given a path \pathCurrent and its features, Algorithm~\ref{alg:scoring} shows the anomaly scoring procedure that \detector uses to make its alerting decision. 
Intuitively, \detector's scoring algorithm generates an alert for \pathCurrent if it has one of the most suspicious feature sets in recent history.

\detector's alerting algorithm, \textsc{AlertGen}, takes three inputs:
a path to score (\pathCurrent), a set of historical paths ($L$) to compute \pathCurrent's anomaly score, and a set of historical alerts ($A$) for paths with unclear causality.
\detector generates the set of historical paths ($L$) by iterating over each login in the historical training data and
running \detector's causality engine to produce an aggregate set of all paths for each login.
For efficiency, \detector can compute this set of historical paths as a batch job at the beginning of each week, and reuse it for the entire week's scoring.
The historical set of alerts ($A$) consists of the \budgetSym$\times$ $H$ most suspicious paths during the historical window, where $H$ is the number of days in the historical window and $B$ is the user-provided alert budget.

With these three inputs, \detector computes an anomaly score for \pathCurrent that represents the fraction of historical paths where \pathCurrent had more (or equally) suspicious feature values.
\detector then compares \pathCurrent's anomaly score against the scores of the historical alerts, and generates an alert for \pathCurrent if its score exceeds any alert's score;
\ie~\detector produces an alert if \pathCurrent is at least as suspicious as a previous alert's path.

\paragraph{Computing Scores}
Conceptually, a path \pathCurrent's anomaly score corresponds to a cumulative tail probability: how much more suspicious (unlikely) is \pathCurrent relative to the kinds of paths that benign users historically make?
As illustrated in the \textsc{Score} subroutine in Algorithm~\ref{alg:scoring}, \detector computes this score
by computing a sub-score for each of the path's features, and then multiplies these sub-scores to get an overall score.

Each feature's sub-score estimates the fraction of historical paths where \pathCurrent had a more suspicious feature value.
In practice, imprecision from \detector's path inference algorithm could lead a naive computation of this fraction to over-count certain historical paths.
For example, a historical login from a server with many ($N$) inbound logins will generate $N$ historical paths,
even though only one of those paths reflects a true causal path.
These types of paths, that involve servers with many inbound logins, will have an inflated volume that could skew the anomaly sub-scores that \detector computes; \ie their features will be over-represented in the historical distribution.
To mitigate this problem, when computing the set of paths for each historical login \newLogin, \detector annotates each path with a ``Path Certainty'' fraction,
denoted as $C$, that equals 1 / the total number of causal paths that \detector inferred for \newLogin.
When \detector computes each sub-score for the current path \pathCurrent, it uses $C$ to down-weight the impact of each historical path (Line 5 of the \textsc{Sub-Score} routine in Algorithm~\ref{alg:scoring}).
\paragraph{Alert Clustering}
To avoid generating redundant alerts for the same path, \detector clusters its alerts each day.
\detector maintains a list of every alert (path) it generates on the current day.
If a new alert path traverses the same exact edges as any path on the day's alert list,
\detector updates the existing alert with information about this duplicate path and does not generate a new alert. 

\subsection{Real-time Detection}
Organizations can run \detector as a real-time detector using an architecture similar to the design described above.
For real-time detection, \detector would maintain a ``recent login'' queue of all logins over the past $T$ hours,
where $T$ corresponds to the causality threshold described in \S~\ref{sec:paths}.
For each new login, \detector can run the path inference procedure described in Section~\ref{sec:paths},
and then apply the scoring algorithms described above to determine whether any path produces an alert.
Each night, \detector can prune the queue of recent logins to only retain those in the past $T$ hours,
the set of historical paths used for feature extraction,
and update the set of the historical alert paths that \detector uses when assessing a new path's anomaly score
(Section~\ref{sec:detection:detector_switch_unclear}).
This real-time architecture retains the same detection accuracy as running \detector as a batch detector,
since it makes no difference whether \detector classifies each day's logins individually or in one aggregate batch.

\section{Evaluation}
\label{sec:evaluation}

We evaluated \detector on our \totalMonths-month dataset, measuring its detection rate (fraction of attacks detected) and the volume of false positives it generates.
Our data does not contain any known lateral movement attacks, but it does contain one in-situ lateral movement attack conducted by \dropbox's professional red team.
Additionally, we generated and injected a realistic and diverse set of \numAttackSynthetic simulated attacks into our data for a more thorough evaluation (\S~\ref{sec:evaluation:attack_data}).
\detector successfully detected \tprLM of the attacks in our data, including the red team attack,
while generating an average of \numAvgDailyAlerts false positives per day (\S~\ref{sec:evaluation:results}):
an \fpReduction reduction in the number of false positives produced by prior state-of-the-art (\S~\ref{sec:evaluation:baseline}).

\subsection{Implementation}
\label{appendix:implementation_numbers}

\begin{table}
\centering
\begin{tabular}{rr}
	\toprule
	\textbf{Path} & \textbf{\# of Paths with Potential } \\
	\textbf{Length}& \textbf{Credential Switch}
    \\ \midrule
        2 & 3,357,353 \\
        3 & 829,044 \\
        4 & 128 \\
        5 & 6 \\
        6 & 4
    \\ \bottomrule
\end{tabular}
\setlength{\belowcaptionskip}{-10pt}
	\caption{The volume of multi-hop paths, with a potential switch in credentials, inferred by \detector's causality engine. 
	The left column reports the path length and the right column reports the total number of paths with that length that \detector generated, across our dataset.
	}
	\label{table:long_path_distrib}
\end{table}

For our experiments, we implemented \detector in Python 2.7 on a Linux server with 64GB of RAM and a 16-core processor.
Table~\ref{table:long_path_distrib} shows the total number of multi-hop paths that \detector generated, based on the optimized implementation described in Appendix~\ref{appendix:implementation:paths}.
In aggregate, the full set of paths (containing the attributes described in Table~\ref{table:path:schema} and their feature values) consume a total of 2.5GB of memory.
Running \detector's path generation algorithm across our entire data set took a total CPU time of 35 minutes and 13 seconds, and
running \detector's feature extraction and detection algorithms on every day of in our data set took a cumulative CPU time of 83 minutes and 9 seconds.

The dramatic drop in long-length paths reflects the fairly flat topology of \dropbox's network,
the filtering steps that \detector takes to remove noisy and spurious login activity (\S~\ref{sec:data:cleaning}),
and the optimization \detector uses of only tracking paths with potential (or clear) credential switching.
System administrator activity predominates these multi-hop paths, since most other users perform logins directly into their target service (\eg short one-hop paths).

\subsection{Attack Data}
\label{sec:evaluation:attack_data}

\paragraph{Red Team Attack}
Our data contains one lateral movement attack generated by \dropbox's professional red team.
The red team began their attack from a ``compromised'' employee's laptop (randomly selected from an existing pool of volunteer employees).\footnote{The red team followed their standard safety protocols when conducting this simulation,
which included obtaining prior consent from all ``compromised users'', coordinating extensively with the security incident response team, and conducting any necessary remediation that resulted from the simulated attack
(\eg resetting any credentials that they accessed).}
Their attack simulated a common APT scenario~\cite{ryuk,ukraineLM2}, where an attacker conducts lateral movement to access an organization's Domain Controllers (credential management servers).
From their initial foothold, the red team conducted a series of reconnaissance and internal login (lateral movement) operations.
They identified and acquired a new, elevated set of credentials, which they then used to access one of the organization's Domain Controllers.
Apart from requiring that their movement occurred via logins (as opposed to exploiting a machine vulnerability),
the red team performed this attack under no constraints or input from us.
We did not examine the red team data until we had frozen the design and parameters of our detector.
The red team's attack created an \textsc{Unclear} path,
because the attack ``stole'' and then used a sysadmin's credentials from a server that had a recent inbound login by the sysadmin.
\detector's unclear causality detector successfully identified this attack.
Based on its anomaly score, \detector ranked this attack path as the most suspicious path on that day and the 45th most suspicious path across all paths during the month of the attack.

\paragraph{Realistic Attack Simulations}
\dropbox employs multiple sets of security controls and detection approaches,
including commercial security products, external security audits, and custom tools developed by in-house security teams.
Across all of these sources, no incidents of real-world lateral movement have been detected.
Given the lack of real-world attack instances,
we developed an attack synthesis framework and generated an additional \numAttackSynthetic realistic lateral movement attacks.
Our attack framework covers a wide range of real-world attacks described in public breach reports and academic surveys~\cite{saleem2020sok},
ranging from ransomware to targeted APT attacks.\footnote{Our simulation code is available at \url{https://github.com/grantho/lateral-movement-simulator}}

We randomly selected \totalAttackUsersLM employees in our data as starting victims,
whose machines served as ``compromised'' footholds for attackers to launch their lateral movement.
For each starting victim, our framework synthesized twelve different attack scenarios,
corresponding to a pairing of one of three \textsc{Attack Goals} with one of four types of \textsc{Stealthiness}.

Given a starting victim and attack scenario, our framework synthesizes a set of lateral movement login entries that began at a random date and time (when the starting victim was still active in our data).
Leveraging the global graph of all logins in our dataset, our framework simulates an attacker who
iteratively (1) accrues a set of ``compromised'' credentials (the starting victim's credentials, and after each new login, the users who recently accessed the login's destination machine),
and then (2) synthesizes login entries to new destinations that their compromised credential set could access.

The three attack goals specify when an attack succeeds (stops generating new logins) and the shape of the attack's movement.
Modeling ransomware, an \textit{Aggressive Spread} attack generates new logins by iterating over its compromised credential set and performs logins into every machine accessible by each credential; this attack terminates after accessing 50 machines, or once it makes a login into every machine available to its final credential set.
An \textit{Exploratory Attack} stops generating new logins once it accesses a machine that its initial victim did not have access to; this attack iteratively generates new logins by randomly selecting a credential from its compromised set and a new destination accessible to the selected credentials.
\textit{Targeted Attacks} perform logins until they access a high-value server (\eg Domain Controllers).
These attacks generate logins by computing a shortest path to elevated credentials that can access a high-value server, and then compute a shortest path that uses these new credentials to access the high-value server.

Additionally, our attack framework only produces logins that follow the scenario's specified stealthiness.
An attack with \textit{Prior Edge} stealthiness only generates logins that traverse edges that legitimate users had previously made. 
An attack with \textit{Active Credential} stealthiness only uses a set of credentials in a login if the credential's legitimate user was recently logged into the source machine (\ie creating login paths with unclear causality).
An attack with \textit{Combined Stealthiness} only generates logins with both of the properties above (\eg mimicry-style attacks).
The fourth type corresponds to an attacker without any stealthiness requirements.

We generated 326 successful attacks, with 205 attacks across the three stealthier levels (Table~\ref{table:evaluation:results});
users did not always have viable attack paths, leading to less than 50 attacks per scenario (\eg users with limited access or who lacked stealthy paths for a targeted attack).
The red team attack corresponded to a \textit{Targeted Attack} with \textit{Active Credential} stealthiness;
our framework can produce the same attack path if we run it from the same starting victim with these parameters.


\begin{table}
\centering
\resizebox{\columnwidth}{!}{
\begin{tabular}{lrrr|r}
  \toprule
\multicolumn{2}{r}{\textbf{Exploratory}} & \textbf{Aggressive} & \textbf{Targeted}  & \textbf{TP Rate} \\
	 \midrule
    \textbf{No stealth}$\dagger$ & 37 / 41 & 38 / 41 & 38 / 40 & 113 / 122 \\
    \textbf{Prior Edge} & 13 / 14 & 14 / 14 & 10 / 13 & 37 / 41 \\
    \textbf{Active Cred.} & 41 / 41 & 41 / 41 & *39 / 41 & 121 / 123 \\
    \textbf{Combined} & 12 / 14 & 14 / 14 & 12 / 13 & 38 / 41 \\
\midrule\addlinespace
\textbf{Detection Rate} & 103 / 110 & 107 / 110 & 99 / 107 & 309 / 327 \\
\end{tabular}
}
\setlength{\belowcaptionskip}{-10pt}
	\caption{Summary of \detector's detection (true positive) rate across the different scenarios simulated by our attack framework and the red team attack (\S~\ref{sec:evaluation:attack_data}).
	Rows correspond to the four different stealthiness levels and columns correspond to the three attack goals that our framework simulated for each user.
	The last column and last row report \detector's overall detection (TP) rate.
	The scenario marked with an asterisk (\textsc{Targeted} and \textsc{Active Cred}) includes one red team attack, which \detector detected.
	$\dagger$The false negatives in the ``No stealth'' row stem from inaccurate attributes in the attack logins.
	}
  \label{table:evaluation:results}
\end{table}

\subsection{Results}
\label{sec:evaluation:results}

\paragraph{Evaluation Procedure}
We divided our data into a 2-month training window (Jan 1 -- Mar 1, 2019), 
which we used to bootstrap the feature extraction and scoring components of \detector that require historical data,
and a \numEvalMonths-month evaluation window (\dateEvalStart to \dateEnd).
Our evaluation data contained \numEvalLogins successful logins, and \numFinalEvalLogins logins after applying \detector's data filtering steps (\S~\ref{sec:data:login_filtering}).
We ran \detector over this evaluation data to compute its false positive rate and detection (true positive) rate.
For any detection component that required historical training data, we used a rolling window of the preceding \trainingWindowDays days.
For our anomaly scoring algorithm (\S~\ref{sec:detection:detector_switch_unclear}), we used a budget of \initialAlertBudget alerts / day, and explore the sensitivity of this parameter below.

\paragraph{Attack Detection Rate (True Positives)}
For each of the \numAttackSynthetic attacks synthesized by our framework,
we injected the attack's logins into our evaluation data and ran \detector on the day(s) when the attack occurred.
For the red team exercise, we examined the alerts that \detector generated on the day of the attack.
We deemed \detector successful if it generated an alert for any attack path made by the simulated attacker or red team.

Table~\ref{table:evaluation:results} shows that \detector successfully detected
a total of \numDetectedTotal attacks (\tprLM),
which includes the attack performed by \dropbox's expert red team.
\detector detected \numDetectedClearSwitch attacks through its rule set for paths with clear credential switching (\S~\ref{sec:detection:detector_switch_clear}).
In all of these attacks, the simulated attacker either used a new set of credentials in a login from their initial foothold machine or from a server that the legitimate user (of the new credentials) had not recently accessed,
enabling \detector to identify a movement path where the attacker clearly switched to using new credentials.

However, most (\numAttackUnclearSwitchSynthetic) attacks created paths with \textsc{Unclear} causality,
either because the attack quickly capitalized on new credentials that were recently used on a server, or because the attack simulated a stealthy adversary who only used new credentials from machines where the legitimate user was recently or currently active.
Detecting these paths falls to \detector's anomaly scoring detector (\S~\ref{sec:detection:detector_switch_unclear}).
With a budget of \initialAlertBudget alerts per day,
\detector successfully identified \numDetectedUnclearSwitch of these attacks (95\%),
including the red team attack. 

\paragraph{False Negatives}
Of the \numFNLM false negatives,
\detector missed \numFNBrokenEnrichmentLM attacks because of attribute errors in the login data. 
For each of these \numFNBrokenEnrichmentLM false negatives,
the attack logins had an incorrect client vs. server label for a machine,
and/or contained incorrect information about a machine's owner.
If we replaced this inaccurate login information with the correct attributes
(acquired from additional, up-to-date data sources at \dropbox),
\detector could successfully detect all \numFNBrokenEnrichmentLM of these false negatives with its \textit{clear credential switch} detector.
Nonetheless, we count these attacks as false negatives since real data inevitably contains imprecise information.
Additionally, \detector failed to detect \numFNLowBudgetLM stealthy attacks using a daily budget of \initialAlertBudget alerts.
For all of these false negatives, every attack login traversed an edge with at least three prior days where the legitimate user had performed a login along the edge.

\begin{figure}[t]
\centering
\includegraphics[width=0.99\columnwidth]{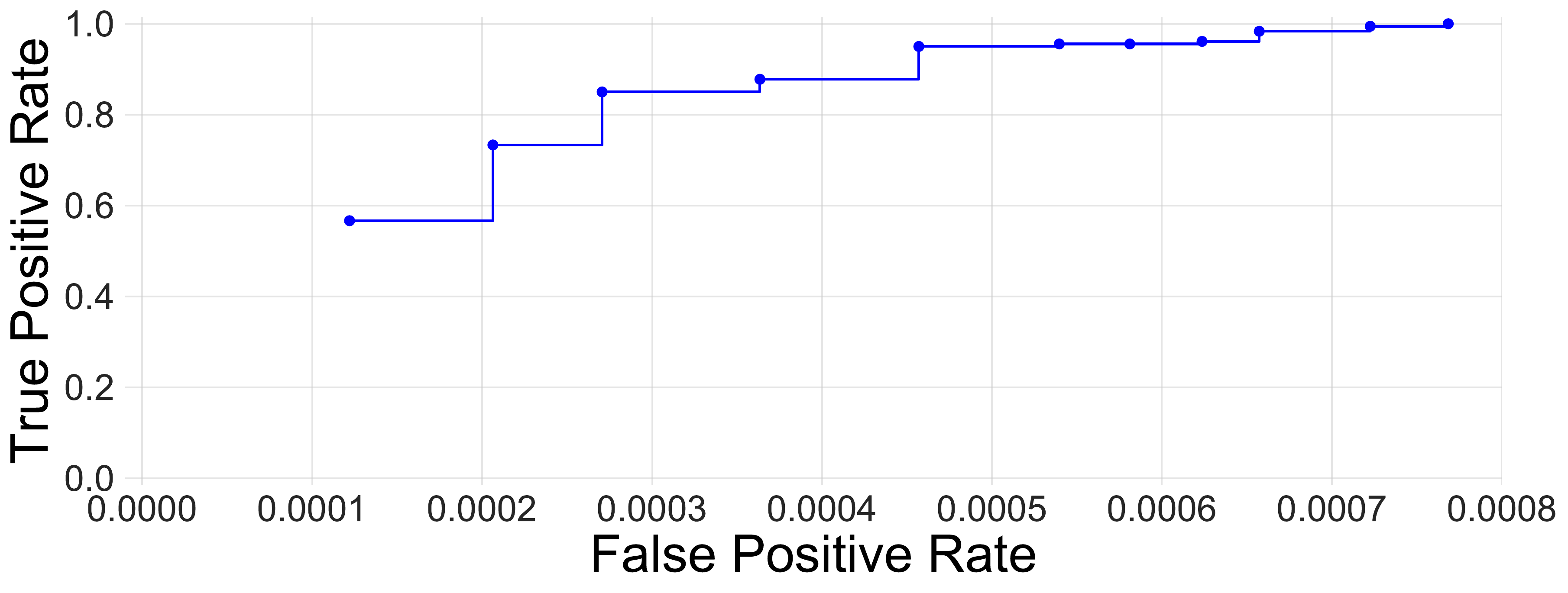}
\caption{
ROC Curve for \detector's unclear causality detector (\S~\ref{sec:detection:detector_switch_unclear}) at different budgets (1--11 daily alerts).
The True Positive Rate reports the fraction of (\numAttackUnclearSwitchSynthetic) attacks with unclear causality that \detector detects.
The FP Rate reports the number of false alarms divided by the number of logins in our evaluation data (2.94M).
}
\label{fig:hopper_roc}
\end{figure}

\begin{figure}[t]
\centering
\includegraphics[width=0.99\columnwidth]{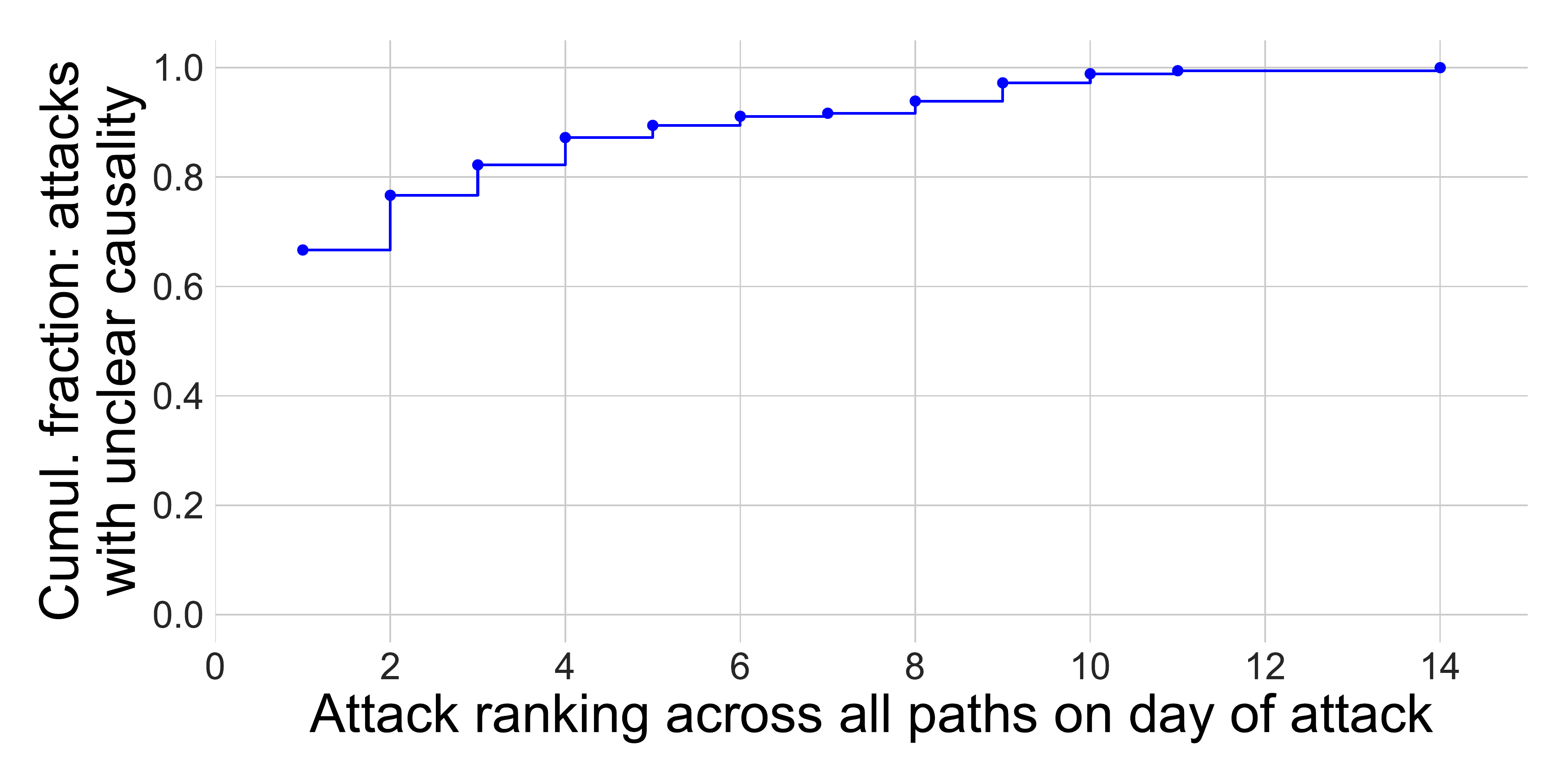}
\setlength{\belowcaptionskip}{-10pt}
\caption{
The ranking of attack paths with \textsc{Unclear} causality,
relative to all of the login paths that occurred on the day of an attack.
}
\label{fig:attack_unclear_ranking_daily}
\end{figure}

\paragraph{Budget Sensitivity and Attack Rankings}
Including the red team attack, \numAttackUnclearSwitchSynthetic attacks produced paths with unclear causality.
Figure~\ref{fig:hopper_roc} shows the detection performance of \detector for these attacks, using different daily budgets for its anomaly scoring detector.
\detector uses this budget to build a set of the historical alerts over the past month, and then alerts on a new path (with unclear causality) if its score is greater than or equal to any scores of the historical alerts (\S~\ref{sec:detection:detector_switch_unclear}).
If \detector used a daily budget of 11 alerts, it could eliminate \numFNLowBudgetLM false negatives and detect all \numAttackUnclearSwitchSynthetic attacks with a false positive rate of 0.00076.

We also assessed the ranking of these \textsc{Unclear Path} attacks relative to the benign paths in our data, based on their anomaly scores. 
Figure~\ref{fig:attack_unclear_ranking_daily} shows that
\detector ranks these attacks as highly suspicious, with over 66\% of attacks ranked as the most suspicious path on the day each attack occurred.

\paragraph{False Positives}
To compute \detector's false positive rate,
we ran \detector on all non-synthesized logins for each day in our evaluation data.
We conservatively labeled all of the alerts \detector produced as false positives if they did not relate to the red team attack.

With a daily budget of \initialAlertBudget alerts for its anomaly scoring detector,
\detector's two detection algorithms generated a total of \numTotalAlerts false positives (FP) across the \numEvalDays-day evaluation window:
an average of \numAvgDailyAlerts alerts / day and a false positive rate of \fpr across the \numFinalEvalLoginsApprox filtered logins in our evaluation data.
\detector's rule-based detector for \textsc{Clear} paths produced \numAlertsClearSwitching FP's,
and the remaining \numAlertsUnclearSwitching FP's come from \detector's anomaly scoring detector.
On some days, \detector's anomaly scoring detector generated less than \initialAlertBudget alerts because (1) not every day had \initialAlertBudget suspicious paths with unclear causality (\eg weekends and holidays), and (2) our alert clustering resulted in some days with fewer alerts (\S~\ref{sec:detection:detector_switch_unclear}).

We identified several common reasons for many of these false positives.
Across the \numAlertsClearSwitching false positives generated by our \textsc{Clear} path detector,
approximately 10\% of these false positives correspond to logins where a user's laptop accesses a particular service using a special service account.
Another 41.5\% correspond to machine imaging and provisioning activity,
where a sysadmin runs a script that uses their elevated set of credentials to configure a laptop for a new owner
(these logins occurred at a remote office that \detector's data cleaning steps did not filter out).
Finally imprecision in \detector's causality engine contributed to 19\% of \detector's \textsc{Clear} path false positives and over 49\% of \detector's \textsc{Unclear}-causality false positives.
Many of these false positives are paths, initiated by one system administrator, that purportedly make a login that switches to another system administrator's credentials.
These alerts often involve a handful of ``gateway'' machines that sysadmins use to access important internal servers (\eg Domain Controllers).
\detector generates these false alerts when multiple sysadmins have recently logged into a gateway machine,
and one sysadmin launches a login from the gateway machine to a rarely-accessed or niche server.
Because these paths involve only administrator credentials, \detector could reduce its false positives by filtering them out; any credential switch between two administrators likely provides limited additional access.

\begin{table}
\small
\centering
\resizebox{\columnwidth}{!}{
\begin{tabular}{lcr}
	\toprule
	\textbf{Detector} & \textbf{Detection Rate} & \textbf{False Positives} \\
    \midrule
    SAL (equal FP) &
        \detectionRateHalf (47.7\%) & \ccsBaselineNumAlertsHalf (0.12\%) \\
    SAL (equal TP) &
        \detectionRate (94.5\%) & \ccsBaselineNumAlertsFull (0.94\%) \\
    \midrule
    \detector & \detectionRate (94.5\%) & \numTotalAlerts (0.12\%)
    \\ \bottomrule
\end{tabular}
}
\setlength{\belowcaptionskip}{-10pt}
	\caption{
	Prior state-of-the-art, \ccsDetector~\cite{ccs2017lmdetecting},
	produces \fpReduction as many FP as \detector to detect the same number of attacks.
	At a similar number of FP's as \detector, \ccsDetector detects roughly half as many attacks (\S~\ref{sec:evaluation:baseline}).
        }
	\label{table:eval:baseline}
\end{table}

\subsection{Comparison with Prior State-of-the-Art}
\label{sec:evaluation:baseline}

We compared \detector's performance against the best performing prior work,
the Structurally Anomalous Login (\ccsDetector) detector proposed by Siadati and Memon~\cite{ccs2017lmdetecting}.
\ccsDetector detects lateral movement by generating a set of logins that traverse a rare edge in the login graph (based on a user-specified threshold).
Next, \ccsDetector learns and uses a set of ``benign login patterns'' to identify which rare edges to alert on.
Each login pattern corresponds to a triplet of (source machine attributes, destination machine attributes, and user attributes).
For example, given the login (src = Machine A, dest = Machine B, user = Alice),
(src = New York, dest = San Francisco, user = Engineering) would be one login pattern, if Machine A resides within New York, Machine B resides within San Francisco, and Alice works on the Engineering team.
\ccsDetector learns a set of benign patterns by using a historical set of logins to identify patterns where a sufficiently large fraction of source machines, destination machines, and/or users have at least one historical login that matches a pattern.
\ccsDetector then produces an alert for every rare-edge login that does not match a benign pattern.

Based on the data available to us, we use the following set of login attributes from the SAL paper:
each user has two attributes: (the user's team ,and the user's type: system administrator, regular user, or service account)
and each machine has two attributes: (the machine's type: client or service, and the machine's geographic location).
We applied SAL with a rolling two-month training window on all of the filtered logins in our evaluation window
(\ie the same data used for \detector's evaluation; we also applied both the data filtering and benign scenario pruning outlined in~\S~\ref{sec:data:login_filtering} and~\S~\ref{sec:detector}).
\ccsDetector takes two user-provided thresholds for training and classification, respectively.
Table~\ref{table:eval:baseline} reports the results for SAL using the parameters that produced the minimum volume of FP's to detect (1) the same number of attacks as \detector and (2) (approximately) half as many attacks as \detector.
We report the number of FP's \ccsDetector produces after de-duplicating the alerts to only include one edge (source, destination, and target user) per day,
and we considered \ccsDetector successful if it produced an alert for \textit{any} malicious login in an attack.

\ccsDetector produces nearly \fpReduction as many false positives as \detector to detect the same number of attacks.
Whereas \detector's selectively chooses when to apply anomaly detection (to resolve uncertainty in paths that might have the two key attack properties), \ccsDetector follows a traditional machine learning approach by
simply applying anomaly detection to every login, resulting in significantly more false positives.

\subsection{Attack Case Studies}
\label{sec:evaluation:case_studies}
Below, we describe two attacks created by our synthesis framework, and examine how \detector and
traditional anomaly detection approaches, such as \ccsDetector, handle them.

\paragraph{Example Attack 1: Targeted Compromise}
One attack simulated an adversary who began their lateral movement from an engineer's laptop and then
attempted to access one of several high-value machines within an organization
(\eg a Domain Controller).
After three logins, the attacker arrived on a machine where a system administrator,
\switchingLoginUser, had recently logged into the machine via ssh.
Simulating an attacker compromising and using Bob's ssh credentials (\eg by abusing a forwarded SSH agent),
our framework created a fourth attack login that leveraged Bob's credentials to access
a server that manages user permissions and SSH keys.

The last two logins involved in this attack path rarely occur, enabling \ccsDetector to detect this
attack with a low volume of false positives.
Similarly, \detector successfully detects this attack, even though it involves an attack path with
unclear causality (since the sysadmin had an active ssh session that could have
launched the final login into the ssh management server); the rareness of
the attack path's edges led \detector to rank it among the top 10 most suspicious paths that month.

\paragraph{Example Attack 2: Stealthy, Short Paths}
For each user, our framework also simulated attacks that modeled a stealthy
adversary who only accesses machines via previously traversed graph edges.
In one such attack, starting from a compromised user (\causalUser)'s machine,
our framework first synthesized a login to a server (\switchingSrc) that \causalUser had previously accessed
(4 out of the past 60 days).
After moving to Server \switchingSrc,
the attacker observed that Server \switchingSrc still had the credentials of a sysadmin, \switchingLoginUser,
cached from a login during the past week,
enabling the attacker to acquire them.
The attacker (our framework) also observed that
\switchingLoginUser had previously logged into a powerful remote management
machine from Server \switchingSrc (3 out of the past 60 days).
Accordingly, our framework synthesized a final, second attack login using
\switchingLoginUser's credentials to access this high-value server.
Although seemingly simple, this attack reflects a realistic path for a
stealthy attacker, since shorter paths provide fewer opportunities for detection. 

\detector detected this attack
with its \textsc{Clear} path detector: the second login
switched to a new target username, but over 24 hours elapsed since
\switchingLoginUser accessed Server \switchingSrc.  Even if
\switchingLoginUser had logged into Server \switchingSrc more recently,
\detector would have caught this attack under its
anomaly scoring detector (which ranks the attack path among the top 20
most suspicious in the past month).
In contrast, because this attack only traverses edges with prior history,
\ccsDetector would produce over 14,000 alerts across our \numEvalMonths-month evaluation data to detect it.

\section{Discussion}
\label{sec:discussion}
\detector achieves good results on the real-world data set we used.
However, a number of interesting future directions remain,
including overcoming potential evasion strategies,
understanding how \detector generalizes across different enterprise network architectures,
and extending \detector's detection approach to achieve better performance.

\subsection{Evasion and Limitations}
An attacker might evade detection if they can piggyback on top of a series of logins made by legitimate users to access their target machines~\cite{niakanlahijishadowmove}, or if the attacker finds a frequently traveled login path that provides access to their target machines.
Our evaluation explicitly generated attacks that pursued this stealthy strategy,
and \detector could detect many of these attacks.
The attacks that \detector failed to detect had \textsc{Unclear} causality,
followed paths with frequently traveled edges, and occurred on days with other \textsc{Unclear} paths whose edges occurred more infrequently.
However, we note that attackers might not always be able to make such stealthy movement:
when synthesizing attacks across our sample of 50 random starting users,
37 users could not stealthily access a high-value server;
\ie attackers who compromised these users' machines had no path to these sensitive machines, or required them to make at least one rare-edge login.

Although our threat model focuses on interactive attackers who manually perform their movement,
attackers could evade detection by installing stealthy malware on a shared server
that lies on the path to their final target machine.
Such malware could wait until the maximum session duration (time threshold for causally linking two logins together) has elapsed.
Once this time has elapsed, the malware could then opportunistically launch the subsequent logins in its attack path
whenever a legitimate user (\eg \textit{Bob}) performs an inbound login into the shared server.
This strategy will cause \detector to causally link the second half of the attack path, that abuses Bob's credentials, to Bob's earlier legitimate logins, creating a \textsc{Benign} path that appears to consistently use one set of credentials.
Because this approach increases attacker dwell time and their host footprint,
complimentary techniques such as binary allow-listing, anti-virus, and additional detection signals (\S~\ref{sec:improvements}) can help increase the chance of detection.

Missing or inaccurate logging information can also create false negatives,
a problem common to any detection strategy.
Future work can explore ways to alleviate this challenge by using multiple sources of information to determine the correct attributes of login data.
Additionally, organizations can deploy commercial log-hygiene solutions to continuously monitor and collate their logging data.

\subsection{Generalizability} 
\label{sec:generalizability}
Although we evaluate \detector on a large real-world data set, \detector's performance could change at
enterprises with significantly different network architectures and security policies.
For example, \dropbox makes a dedicated effort to scope employee access based on the least privileges principle;
at organizations where many users have highly privileged access, an attacker may not need to acquire additional credentials to achieve their desired goal.
As a result, lateral movement attack paths might not exhibit a switch in credentials,
allowing adversaries to evade detection.
For such organizations, implementing better permissions hygiene will likely yield greater security benefits than any detection strategy.
We view \detector as a promising direction for securing enterprises against attacks that could succeed
in spite of the adoption of such security best practices.

With respect to the impact of a network's architecture on \detector's performance,
we observe that two properties contribute to \detector’s success:
a relatively flat network topology and consistent workflows across most users that only access a small subset of machines.
Below, we characterize the graph topology at \dropbox, and explain why we believe many organizations will also exhibit these two properties, allowing \detector to generalize to other networks.

\begin{figure}[t]
\centering
\includegraphics[width=0.48\textwidth]{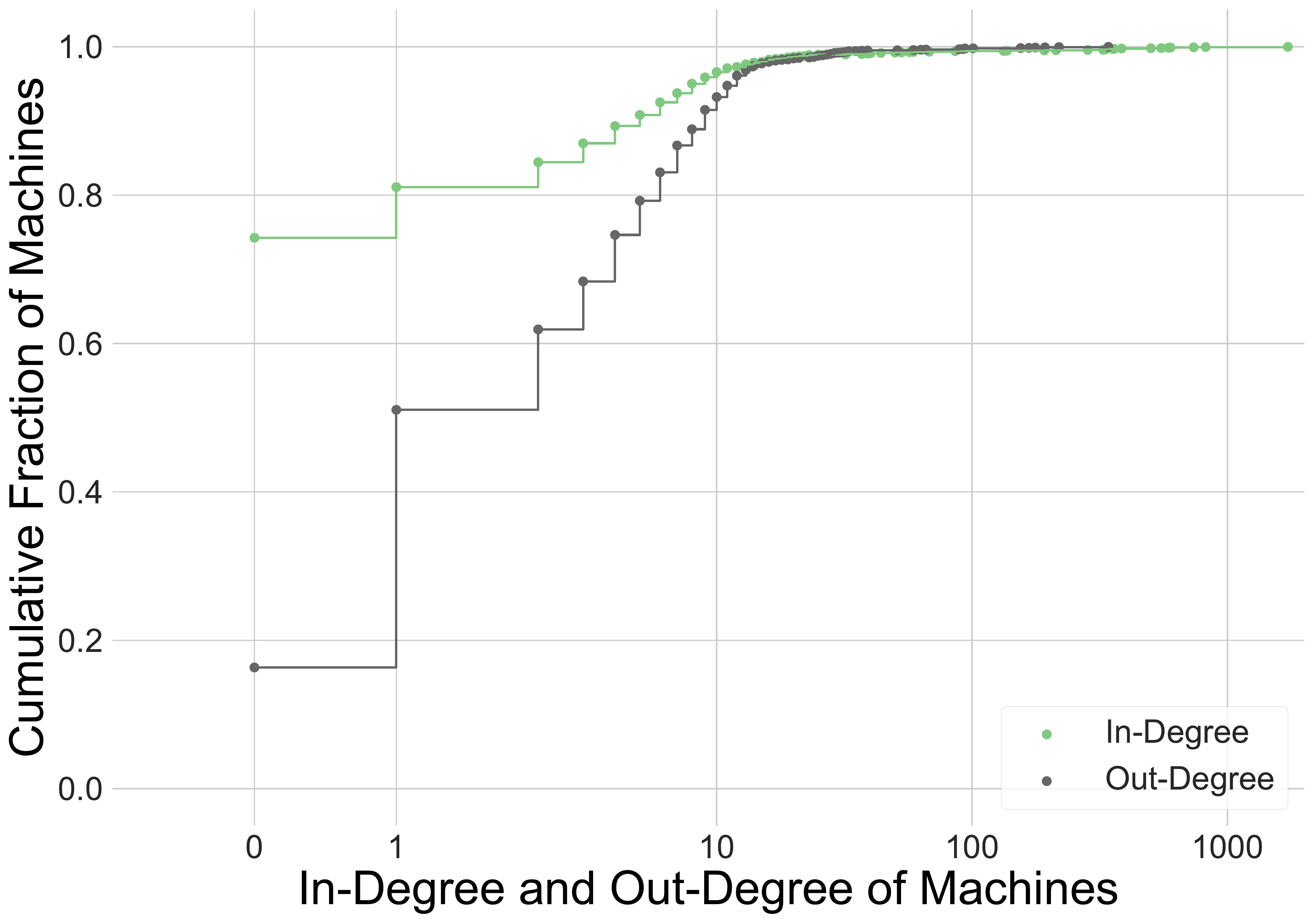}
\setlength{\belowcaptionskip}{-8pt}
\caption{The in-degree and out-degree distribution across hosts at \dropbox.
The in-degree for a host equals the number of machines that it has received logins from;
the out-degree counts how many unique machines each source machine makes at least 1 login into.
}
\label{fig:dbx_degrees}
\end{figure}

\paragraph{Network Topology of \dropbox}
If we aggregate all of the logins across our dataset,
the unified graph has a diameter of length 7 and an average shortest path length of 2.12 hops.
The graph contains 10,434 unique edges, where each edge consists of a (source machine, destination machine) tuple;
when edges also include the username involved in a login, the graph contains 27,718 unique edges.
Figure~\ref{fig:dbx_degrees} shows the in-degree and out-degree distribution for all machines at \dropbox:
\ie the number of distinct machines that a node receives logins from and makes logins to.
The servers with in-degrees of over 100 inbound machines correspond to common enterprise services,
such as Windows Domain Controllers that handle Kerberos-based authentication,
printers, telemetry and logging machines,
and servers involved in provisioning new machines.
Clients (\eg laptops) represent 65\% of the machines in our data,
resulting in many machines with an in-degree of 0.
Machines with high out-degrees (logins to over 100 different destinations)
correspond to system administrator machines, as well as internal scanning and monitoring servers.

\paragraph{Impact of Different Network Configurations}
One of the biggest challenges that \detector faces is the risk of path explosion
and an overwhelming number of suspicious paths with unclear causality.
This situation can occur if many servers have large numbers of users that access them,
who then launch outbound logins from the common servers to other machines;
if this behavior occurs multiple times along a path, it risks an exponential increase in the number of paths that
\detector will infer. 
This path explosion might lead not only to unsuitable run-time performance (\eg consuming too much memory),
but could also lead to a large number of false positives.
If many of these incorrectly inferred movement paths have a suspicious set of features,
then \detector may generate a substantial number of false alerts related to these paths.
Two factors mitigated the problem of path explosion in our data set:
a relatively flat network topology and the natural clustering of user accesses patterns to a few work-related machines.

Flat networks arise because most (non-sysadmin) user activity consists of direct logins from their client machines to the server that hosts their desired functionality or data.
Moreover, because many servers provide a limited UI and set of functionality, they often do not provide an easy way to launch outbound logins.
This property means that even when a server has many inbound logins from users,
it often does not risk path explosion because subsequent outbound logins do not occur.
We expect that even as the number of users and servers increases, these natural habits will keep access patterns relatively flat;
this behavior will increase the number of short login paths, but continue to limit the number of long paths.
At \dropbox, we did observe processes that generated long paths, such as when users need to access a server by tunneling through a gateway (bastion) machine,
automated activity (e.g., domain controllers iteratively synchronizing data amongst each other),
and system administrator activity. 
However, most of the paths from these activities either do not contain both attack properties (\eg no switch in credentials or no new access for the path's potential causal users), or they get removed by \detector's filtering procedure since they do not pose a large risk for lateral movement (\S~\ref{sec:data:login_filtering}).

Second, users tend to access machines for a specific job function,
creating a sparse graph where different subsets of logins naturally cluster around a small group of machines
(\eg at \dropbox over 90\% of machines have an in-degree $\leq 10$ and an out-degree $\leq 10$).
Implementing least privileges, where users have access to only a small set of machines relevant to their work,
also reinforces this common behavior.
As a result, most machines only get accessed by a limited set of users,
which reduces path explosion and the number of paths with unclear causality.
Furthermore, because users accessing a shared server typically work on the same team or have similar job roles,
their credentials often have similar privileges and they tend to access the same broader set of machines.
Thus, even when \detector produces paths with unclear causality,
these paths often do not provide access to an unauthorized machine for their causal user (the second attack property),
and get marked as benign. 
Since this property arises from common user behavior and security policies,
and has been observed at different organizations~\cite{ccs2017lmdetecting},
we expect many other networks exhibit similar partitioning.

\paragraph{\detector's Causality Time Threshold}
\detector uses a time-based threshold, equal to the maximum remote session duration at an organization,
to help infer when logins form a movement path (\S~\ref{sec:paths}).
We discussed this session duration with the security teams of multiple companies,
and all of them implement a similar length policy for remote login sessions (\eg ssh and RDP),
based on commonly-adopted, best-practice recommendations~\cite{nist80063B}, and in some cases compliance and cyber-insurance guidelines~\cite{hipaaTechnicalSafeGuards, irsSafeGuards, pciGuidelines}.
Additionally, even if we doubled the 24-hour threshold that \detector used in our evaluation,
\detector achieves an 89.9\% detection (true positive) rate while generating an average of 9 false alarms / day.

\subsection{Extending Hopper}
\label{sec:improvements}
To further improve \detector's performance,
future work could explore prioritizing paths that involve particularly sensitive credentials or machines.
For example, given a list of sensitive machines,
\detector could assign a higher anomaly score to any path that accesses one of these machines.
Similarly, organizations could tune \detector to focus on paths that involve a credential switch where
the causal user elevates themselves to an administrator account over the course of their logins.

Complementary work uses system logs to detect suspicious host activity that aligns with attacker behavior enumerated in the MITRE ATT\&CK framework~\cite{hassan2020tactical,hossain2020combating,hossain2018dependence,pasquier2017practical}.
Organizations could combine these approaches with \detector to gain insight into both malicious host activity as well as suspicious (lateral) movements between hosts.

Finally, \detector would generate fewer false positives
if it more precisely inferred causally-linked logins.
Future work could explore how drawing upon additional data sets, such as network traffic or host logs,
would enable \detector to infer causality more accurately.
For example, when determining which inbound login causes an outbound login,
\detector could analyze the inbound versus outbound network flows across the
candidate logins to pinpoint pairs with overlapping timing and flow sizes.

\section{Conclusion}
\label{sec:conclusion}
This paper presented \detector, a system that develops a graphical model of enterprise logins to detect lateral movement.
On a \totalMonths-month enterprise data set, \detector detected \tprLM of realistic attack scenarios at a false positive rate of \fpr.
These results illustrate the power of a causal understanding of the movement paths that users make between \textit{internal} enterprise machines.
By understanding which logins belong to the same logical movement path and the user responsible for initiating each path, \detector can identify a diverse range of attacks while generating \fpReduction as many false positives as prior state-of-the-art.
Although common authentication logs make inferring precise causality difficult,
\detector's use of specification-based anomaly detection ---
selectively applying anomaly detection only in cases of high uncertainty ---
enables it to achieve good detection performance.

\section*{Acknowledgements}
We thank Dropbox's security team for supporting this research, and 
John Cramb in particular for his help conducting the red team exercise.
This work was supported in part by the Hewlett Foundation through the
Center for Long-Term Cybersecurity, NSF grants CNS-1237265 and CNS-1705050,
an NSF GRFP Fellowship, the UCSD CSE Postdoctoral Fellows program,
the Irwin Mark and Joan Klein Jacobs Chair in Information and Computer Science
(UCSD), by generous gifts from Google and Facebook, and operational
support from the UCSD Center for Networked Systems.

\bibliographystyle{plain}
\bibliography{references}

\begin{thebibliography}{10}

\bibitem{remctl}
Russ Allbery.
\newblock remctl: Remote authenticated command execution.
\newblock \url{https://github.com/rra/remctl}, 2018.

\bibitem{bohara2017unsupervised}
Atul Bohara, Mohammad~A Noureddine, Ahmed Fawaz, and William~H Sanders.
\newblock An unsupervised multi-detector approach for identifying malicious
  lateral movement.
\newblock In {\em IEEE Symposium on Reliable Distributed Systems (SRDS)}, 2017.

\bibitem{usenix_ti}
Xander Bouwman, Harm Griffioen, Jelle Egbers, Christian Doerr, Bram Klievink,
  and Michel van Eeten.
\newblock A different cup of {TI}? the added value of commercial threat
  intelligence.
\newblock In {\em {USENIX} Security Symposium}, 2020.

\bibitem{lmUnsupervisedGraphAI}
Benjamin Bowman, Craig Laprade, Yuede Ji, and H.~Howie Huang.
\newblock Detecting lateral movement in enterprise computer networks with
  unsupervised graph {AI}.
\newblock In {\em International Symposium on Research in Attacks, Intrusions
  and Defenses ({RAID})}, 2020.

\bibitem{lateralMovementUSCert}
CERT.
\newblock Advanced persistent threat activity targeting energy and other
  critical infrastructure sectors.
\newblock \url{https://www.us-cert.gov/ncas/alerts/TA17-293A}, 2017.

\bibitem{ciscoNetworkSegmentation}
Cisco.
\newblock What is network segmentation?
\newblock
  \url{https://www.cisco.com/c/en/us/products/security/what-is-network-segmentation.html},
  2019.

\bibitem{pciGuidelines}
PCI Security~Standards Council.
\newblock {PCI DSS Prioritized Approach for PCI DSS 3.2}.
\newblock
  \url{https://www.pcisecuritystandards.org/documents/Prioritized-Approach-for-PCI_DSS-v3_2.pdf?agreement=true&time=1469037392985},
  2016.
\newblock Section 8.1.8.

\bibitem{crowdstrikelateralmovement}
CrowdStrike.
\newblock Lateral movement.
\newblock \url{https://www.crowdstrike.com/epp-101/lateral-movement/}, Sep
  2019.

\bibitem{cobaltKitty}
Assaf Dahan.
\newblock Operation cobalt kitty.
\newblock \url{https://www.cybereason.com/blog/operation-cobalt-kitty-apt},
  2017.

\bibitem{dunagan2009heat}
John Dunagan, Alice~X Zheng, and Daniel~R Simon.
\newblock Heat-ray: combating identity snowball attacks using machinelearning,
  combinatorial optimization and attack graphs.
\newblock In {\em ACM Symposium on Operating Systems Principles (SOSP)}, 2009.

\bibitem{targetbreach}
Jim Finkle and Susan Heavey.
\newblock Target says it declined to act on early alert of cyber breach.
\newblock
  \url{http://www.reuters.com/article/us-target-breach-idUSBREA2C14F20140313},
  Mar 2014.

\bibitem{freitas2020d2m}
Scott Freitas, Andrew Wicker, Duen~Horng Chau, and Joshua Neil.
\newblock {D2M: Dynamic Defense and Modeling of Adversarial Movement in
  Networks}.
\newblock In {\em SIAM International Conference on Data Mining}, 2020.

\bibitem{humanOperatedLM2}
Sergiu Gatlan.
\newblock Microsoft shares tactics used in human-operated ransomware attacks.
\newblock
  \url{https://www.bleepingcomputer.com/news/security/microsoft-shares-tactics-used-in-human-operated-ransomware-attacks/},
  Mar 2020.

\bibitem{nist80063B}
Paul~A. Grassi, Elaine~M. Newton, Ray~A. Perlner, Andrew~R. Regenscheid,
  James~L. Fenton, William~E. Burr, Justin~P. Richer, Naomi~B. Lefkovitz,
  Yee-Yin Choong, Kristen~K. Greene, Jamie~M. Danker, and Mary~F. Theofanos.
\newblock {NIST Special Publication 800-63B: Digital Identity Guidelines}.
\newblock \url{https://doi.org/10.6028/NIST.SP.800-63b}, 2017.
\newblock Section 4.3~--~4.5.

\bibitem{anthembreach}
Robert Hackett.
\newblock Anthem, a major health insurer, suffered a massive hack.
\newblock \url{http://fortune.com/2015/02/05/anthem-suffers-hack/}, Feb 2015.

\bibitem{hagberg_connected_2014}
Aric Hagberg, Nathan Lemons, Alex Kent, and Joshua Neil.
\newblock Connected {Components} and {Credential} {Hopping} in {Authentication}
  {Graphs}.
\newblock In {\em {International} {Conference} on {Signal}-{Image} {Technology}
  and {Internet}-{Based} {Systems}}, 2014.

\bibitem{ryuk}
Alexander Hanel.
\newblock Big game hunting with ryuk: Another lucrative targeted ransomware.
\newblock
  \url{https://www.crowdstrike.com/blog/big-game-hunting-with-ryuk-another-lucrative-targeted-ransomware/},
  Jan 2019.

\bibitem{hassan2020tactical}
Wajih~Ul Hassan, Adam Bates, and Daniel Marino.
\newblock Tactical provenance analysis for endpoint detection and response
  systems.
\newblock In {\em IEEE Symposium on Security \& Privacy}, 2020.

\bibitem{hassan2020omegalog}
Wajih~Ul Hassan, Mohammad~A Noureddine, Pubali Datta, and Adam Bates.
\newblock Omegalog: High-fidelity attack investigation via transparent
  multi-layer log analysis.
\newblock In {\em Network and Distributed System Security Symposium}, 2020.

\bibitem{sshLateralMovementIran}
Sarah Hawley, Ben Read, Cristiana Brafman-Kittner, Nalani Fraser, Andrew
  Thompson, Yuri Rozhansky, and Sanaz Yashar.
\newblock {APT39}: An iranian cyber espionage group focused on personal
  information.
\newblock
  \url{https://www.fireeye.com/blog/threat-research/2019/01/apt39-iranian-cyber-espionage-group-focused-on-personal-information.html},
  Jan 2019.

\bibitem{hipaaTechnicalSafeGuards}
HIPAA.
\newblock {HIPAA: 45 CFR § 164.312: Technical safeguards.}
\newblock \url{https://www.law.cornell.edu/cfr/text/45/164.312}, 2013.
\newblock Section 164.312(a)(2)(iii).

\bibitem{ho2017detecting}
Grant Ho, Aashish Sharma, Mobin Javed, Vern Paxson, and David Wagner.
\newblock Detecting credential spearphishing in enterprise settings.
\newblock In {\em {USENIX} Security Symposium}, 2017.

\bibitem{hossain2020combating}
Md~Nahid Hossain, Sanaz Sheikhi, and R~Sekar.
\newblock Combating dependence explosion in forensic analysis using alternative
  tag propagation semantics.
\newblock In {\em IEEE Symposium on Security \& Privacy}, 2020.

\bibitem{hossain2018dependence}
Md~Nahid Hossain, Junao Wang, Ofir Weisse, R~Sekar, Daniel Genkin, Boyuan He,
  Scott~D Stoller, Gan Fang, Frank Piessens, and Evan Downing.
\newblock Dependence-preserving data compaction for scalable forensic analysis.
\newblock In {\em {USENIX} Security Symposium}, 2018.

\bibitem{irsSafeGuards}
IRS.
\newblock {Safeguard Security Report}.
\newblock
  \url{https://www.irs.gov/pub/irs-utl/irs_safeguards_annotated_ssr_template.pdf},
  2014.
\newblock Section 9.3.1.

\bibitem{kent2015authentication}
Alexander~D Kent, Lorie~M Liebrock, and Joshua~C Neil.
\newblock Authentication graphs: Analyzing user behavior within an enterprise
  network.
\newblock {\em Computers \& Security}, 2015.

\bibitem{ukraineLM}
Robert~M. Lee, Michael~J. Assante, and Tim Conway.
\newblock Analysis of the cyber attack on the ukrainian power grid.
\newblock \url{https://ics.sans.org/media/E-ISAC_SANS_Ukraine_DUC_5.pdf}, Mar
  2016.

\bibitem{liu_log2vec_2019}
Fucheng Liu, Yu~Wen, Dongxue Zhang, Xihe Jiang, Xinyu Xing, and Dan Meng.
\newblock Log2vec: {A} {Heterogeneous} {Graph} {Embedding} {Based} {Approach}
  for {Detecting} {Cyber} {Threats} within {Enterprise}.
\newblock In {\em ACM Conference on Computer and Communications Security
  (CCS)}, 2019.

\bibitem{liu2018latte}
Qingyun Liu, Jack~W Stokes, Rob Mead, Tim Burrell, Ian Hellen, John Lambert,
  Andrey Marochko, and Weidong Cui.
\newblock Latte: Large-scale lateral movement detection.
\newblock In {\em IEEE Military Communications Conference (MILCOM)}, 2018.

\bibitem{fireeyeApt1}
Mandiant.
\newblock Apt1: Exposing one of china’s cyber espionage units.
\newblock
  \url{https://www.fireeye.com/content/dam/fireeye-www/services/pdfs/mandiant-apt1-report.pdf},
  2013.

\bibitem{milajerdi2019holmes}
Sadegh~M Milajerdi, Rigel Gjomemo, Birhanu Eshete, R~Sekar, and
  VN~Venkatakrishnan.
\newblock Holmes: real-time apt detection through correlation of suspicious
  information flows.
\newblock In {\em IEEE Symposium on Security \& Privacy}, 2019.

\bibitem{opmbreach}
Ellen Nakashima.
\newblock Chinese breach data of 4 million federal workers.
\newblock
  \url{https://www.washingtonpost.com/world/national-security/chinese-hackers-breach-federal-governments-personnel-office/2015/06/04/889c0e52-0af7-11e5-95fd-d580f1c5d44e_story.html},
  Jun 2015.

\bibitem{lateralMovementNCSC}
NCSC.
\newblock Joint report on publicly available hacking tools.
\newblock
  \url{https://www.ncsc.gov.uk/report/joint-report-on-publicly-available-hacking-tools},
  2018.

\bibitem{niakanlahijishadowmove}
Amirreza Niakanlahiji, Jinpeng Wei, Md~Rabbi Alam, Qingyang Wang, and Bei-Tseng
  Chu.
\newblock Shadowmove: A stealthy lateral movement strategy.
\newblock In {\em {USENIX} Security Symposium}, 2020.

\bibitem{operationsmn}
Novetta.
\newblock {Operation SMN: Axiom Threat Actor Group Report}.
\newblock
  \url{http://www.novetta.com/wp-content/uploads/2014/11/Executive_Summary-Final_1.pdf},
  Nov 2014.

\bibitem{breachCost}
The~Council of~Economic~Advisors.
\newblock The cost of malicious cyber activity to the u.s. economy.
\newblock
  \url{https://www.whitehouse.gov/wp-content/uploads/2018/03/The-Cost-of-Malicious-Cyber-Activity-to-the-U.S.-Economy.pdf},
  Mar 2018.

\bibitem{pasquier2017practical}
Thomas Pasquier, Xueyuan Han, Mark Goldstein, Thomas Moyer, David Eyers, Margo
  Seltzer, and Jean Bacon.
\newblock Practical whole-system provenance capture.
\newblock In {\em Symposium on Cloud Computing}, 2017.

\bibitem{sshLateralMovementChina}
Fred Plan, Nalani Fraser, Jacqueline O’Leary, Vincent Cannon, and Ben Read.
\newblock {APT40}: Examining a china-nexus espionage actor.
\newblock
  \url{https://www.fireeye.com/blog/threat-research/2019/03/apt40-examining-a-china-nexus-espionage-actor.html},
  Mar 2019.

\bibitem{purvine2016graph}
Emilie Purvine, John~R Johnson, and Chaomei Lo.
\newblock A graph-based impact metric for mitigating lateral movement cyber
  attacks.
\newblock In {\em ACM Workshop on Automated Decision Making for Active Cyber
  Defense}, 2016.

\bibitem{doebreach}
Steve Reilly.
\newblock Records: Energy department struck by cyber attacks.
\newblock
  \url{http://www.usatoday.com/story/news/2015/09/09/cyber-attacks-doe-energy/71929786/},
  Sep 2015.

\bibitem{bloodhound}
Andy Robbin, Rohan Vazarkar, and Will Schroeder.
\newblock Bloodhound: Six degrees of domain admin.
\newblock \url{https://bloodhound.readthedocs.io/en/latest/index.html/}, 2020.

\bibitem{saleem2020sok}
Hamza Saleem and Muhammad Naveed.
\newblock {SoK: Anatomy of Data Breaches}.
\newblock {\em Proceedings on Privacy Enhancing Technologies}, 2020.

\bibitem{ccs2017lmdetecting}
Hossein Siadati and Nasir Memon.
\newblock Detecting structurally anomalous logins within enterprise networks.
\newblock In {\em ACM Conference on Computer and Communications Security
  (CCS)}, 2017.

\bibitem{bronzeunion}
Counter Threat Unit~Research Team.
\newblock Bronze union: Cyberespionage persists despite disclosures.
\newblock \url{https://www.secureworks.com/research/bronze-union}, Jun 2017.

\bibitem{mitreattack}
{The MITRE Corporation}.
\newblock {MITRE ATT\&CK Matrix}.
\newblock \url{https://attack.mitre.org/}, 2020.

\bibitem{trendmicrolateralmovement}
TrendMicro.
\newblock Lateral movement: How do threat actors move deeper into your network?
\newblock
  \url{http://about-threats.trendmicro.com/cloud-content/us/ent-primers/pdf/tlp_lateral_movement.pdf},
  2013.

\bibitem{humanOperatedLM1}
Liam Tung.
\newblock Ransomware: These sophisticated attacks are delivering `devastating'
  payloads, warns microsoft.
\newblock
  \url{https://www.zdnet.com/article/ransomware-these-sophisticated-attacks-are-delivering-devastating-payloads-warns-microsoft/},
  Mar 2020.

\bibitem{wikipediaNetworkSegmentation}
Wikipedia.
\newblock Network segmentation.
\newblock \url{https://en.wikipedia.org/wiki/Network_segmentation}, Sep 2019.

\bibitem{wilkens_towards_2019}
Florian Wilkens, Steffen Haas, Dominik Kaaser, Peter Kling, and Mathias
  Fischer.
\newblock Towards {Efficient} {Reconstruction} of {Attacker} {Lateral}
  {Movement}.
\newblock In {\em {Conference} on {Availability}, {Reliability} and {Security}
  ({ARES})}, 2019.

\bibitem{ukraineLM2}
Kim Zetter.
\newblock Inside the cunning, unprecedented hack of ukraine's power grid.
\newblock
  \url{https://www.wired.com/2016/03/inside-cunning-unprecedented-hack-ukraines-power-grid/},
  Mar 2016.

\end{thebibliography}
\appendix

\section{Implementation Optimizations: Path Inference}
\label{appendix:implementation:paths}

The simplified path inference strategy described in Section~\ref{sec:paths} naively applies recursive backward-tracing on each login to infer broader paths of user movement.
In this section, we describe an optimized implementation of \detector's causality engine that
allows \detector to only perform one iteration of backwards-tracing per login (\ie the backwards-tracing terminates after one-hop).
This more efficient implementation of \detector's causality engine uses a two-pass algorithm that only generates long paths (three or more hops) for logins that could plausibly belong to an attack path.

As a starting point, \detector's causality engine takes three inputs:
a new login event (\newLogin) to infer paths for, a set of all recent logins,
and an incrementally updated ``watchlist'' of suspicious login paths.
The first pass of \detector's causality engine aims to make a greedy decision about \newLogin by performing only one round of backwards-tracing;
this first pass produces a set of one-hop or two-hop causal paths, which \detector then passes to its alert generator (\S\ref{sec:detector}).
If \detector's alert generator cannot make a clean benign versus malicious decision,
the causality engine's second pass continues tracking the path and
continually resubmits an updated, longer-length path to the alert generator for each future login in the path (\S\ref{sec:label:multi-hop-watchlist}).

Under this optimized implementation,
if \detector's alert generation stage receives a path that potentially involves a switch in credentials,
it will either generate an alert for it or add the path to its watchlist.
The only paths that get ``fully'' discarded are those that do not involve any possibility for a switch in credentials,
or paths that match one of the benign movement scenarios described in Section~\ref{sec:detector}.

\subsection{First Pass: Constructing Short Causal Paths}
\label{sec:path:basic_switch}
Given a login, the first pass of \detector's causality engine runs a single step of backwards-tracing to generate either a one-hop path (single login) or a set of two-hops paths, depending on whether the login occurs from a client or server.
Let \newLogin represent a login from source machine \srcCurrent to destination machine \dstCurrent with a target user of \userCurrent.

If \srcCurrent is a client, \detector does not perform any backwards-tracing.
Instead, \detector's causality engine outputs a single one-hop path, where the causal user equals the
owner of \srcCurrent;
clients correspond to the start of a user's movement path and logins from these machines should use the credentials of their owner.
If the login's target username does not match the owner of its source machine,
then \detector includes the login in the paths's changepoint list and marks the path as having a \textsc{Clear} switch in credentials.
In the example from Figure~\ref{fig:path_example}, \detector generates three one-hop paths:
one for each of $L_1$, $L_2$, and $L_3$, which all have empty changepoint lists and are thus labeled as \textsc{Benign} paths.

In contrast, if \srcCurrent is a server, \detector produces a set of two-hop paths by running a single iteration of backwards-tracing to infer a set of causal, inbound logins for \newLogin.
\detector then pairs each of these inbound logins with \newLogin to form a set of two-hop paths.
For each path,
\detector sets the path's causal user equal to the target username of the \textit{inbound} hop.
If none of the inferred paths have a causal user that matches the target username in \newLogin,
then \detector labels these paths a \textsc{Clear} credential switches;
otherwise \detector marks them as \textsc{Unclear} paths.

Finally, \detector also takes a user-provided list of special ``bastion'' machines:
hardened gateway servers that provide access to restricted network segments or machines,
and which require users to perform heightened authentication to access these protected parts of the network
(\eg password and hardware-based 2FA authentication during each login).
Whenever \detector encounters a login that originates from a bastion source machine,
it treats this login as the root login for the path and does not perform any additional backwards-tracing prior to this login;
for these paths, \detector treats the username of the bastion login as the path's causal user.
Because bastions require robust forms of authentication,
logins forwarded from bastion source machines (\ie logins that successfully authenticated to the bastion server)
indicate that login's purported username does reflects the true actor responsible for making the login.
In the example from Figure~\ref{fig:path_example}, if Machine Y corresponds to a bastion server,
then given login $L_4$, \detector will not generate or add any two-hop paths to its watchlist.
However, if Machine Y is not a bastion server and Machine A or Machine B are bastion servers, then given login $L_4$,
\detector would still generate three two-hop paths, with initial (root) logins of $L_1, L_2, L_3$ and causal users of Alice, Bob, and Alice respectively.

\begin{table*}[ht]
\centering
\resizebox{\textwidth}{!}{
\begin{tabular}{r|r|r|r|r|r|r|r|r|r|r|r|r|r|r|r}
 & 1 & 2 & 3 & 4 & 5 & 6 & 7 & 8 & 9 & 10 & 11 & 12 & 13 & 14 & 15 \\
 \hline
0.01 & 386 & 596 & 717 & 764 & 808 & 834 & 861 & 888 & 932 & 965 & 990 & 1,009 & 1,026 & 1,039 & 1,058 \\
0.025 & 722 & 1,100 & 1,360 & 1,503 & 1,652 & 1,774 & 1,888 & 1,990 & 2,106 & 2,215 & 2,305 & 2,397 & 2,487 & 2,561 & 2,655 \\
0.1 & 1,809 & 2,729 & 3,416 & 3,979 & 4,415 & 4,793 & 5,132 & 5,463 & 5,792 & 6,096 & 6,377 & 6,667 & 6,976 & 7,245 & 7,528 \\
0.2 & 2,347 & 3,556 & 4,509 & 5,309 & 5,942 & 6,505 & 6,951 & 7,393 & 7,832 & 8,243 & 8,646 & 9,054 & 9,445 & 9,809 & 10,177 \\
0.5 & 8,025 & 11,402 & 13,978 & 15,767 & 17,359 & 18,726 & 19,947 & 21,110 & 22,175 & 23,180 & 24,192 & 25,187 & 26,143 & 27,064 & 27,927 \\
0.75 & 11,972 & 16,813 & 20,508 & 23,191 & 25,494 & 27,558 & 29,365 & 31,034 & 32,554 & 33,977 & 35,395 & 36,779 & 38,019 & 39,259 & 40,406 \\
\end{tabular}
}
	\caption{Baseline Comparison: the number of false positives generated by SAL~\cite{ccs2017lmdetecting} on our evaluation data under different parameter combinations (Appendix~\ref{appendix:baseline}).
	Each column corresponds to a threshold value for an anomalous edge and each row corresponds to a minimum threshold value for the fraction of users or machines that a ``benign login pattern'' must match.
    }
	\label{table:eval:baseline_full_fp_grid}
\end{table*}

\begin{table*}[t]
\centering
\begin{tabular}{r|r|r|r|r|r|r|r|r|r|r|r|r|r|r|r}
 & 1 & 2 & 3 & 4 & 5 & 6 & 7 & 8 & 9 & 10 & 11 & 12 & 13 & 14 & 15 \\
    \hline
0.01 & 30 & 33 & 34 & 34 & 34 & 34 & 34 & 34 & 34 & 34 & 34 & 34 & 34 & 34 & 34 \\
0.025 & 55 & 72 & 73 & 73 & 74 & 76 & 76 & 81 & 81 & 82 & 82 & 83 & 83 & 83 & 83 \\
0.1 & 98 & 126 & 130 & 139 & 150 & 155 & 160 & 169 & 172 & 174 & 178 & 181 & 185 & 185 & 186 \\
0.2 & 117 & 156 & 165 & 175 & 190 & 201 & 206 & 217 & 222 & 225 & 231 & 234 & 239 & 240 & 244 \\
0.5 & 188 & 231 & 245 & 251 & 260 & 274 & 275 & 288 & 298 & 302 & 303 & 305 & 306 & 307 & 309 \\
0.75 & 222 & 256 & 267 & 271 & 279 & 289 & 291 & 304 & 311 & 315 & 317 & 318 & 319 & 320 & 320 \\
\end{tabular}
	\caption{Baseline Comparison: the number of attacks that SAL~\cite{ccs2017lmdetecting} detects under different parameter combinations (Appendix~\ref{appendix:baseline}).
    }
	\label{table:eval:baseline_full_tp_grid}
\end{table*}

\subsection{Second Pass: Tracking Long Causal Paths}
\label{sec:label:multi-hop-watchlist}

The second pass of \detector's causality engine allows \detector to trace and construct paths of arbitrary length (three or more hops).
To do so, \detector maintains a \textit{watchlist} of earlier paths that \detector's detection algorithm could not clearly label as benign or suspicious:
specifically, paths that potentially involve a switch in credentials (\ie \textsc{Unclear} or \textsc{Clear} paths), do not match an explicit benign scenario, and do not generate an alert via \detector's two detectors).
Initially, this watchlist starts off empty.
As \detector processes and infers causal paths for new logins,
it incrementally adds some of these paths to the watchlist if \detector's alert generator cannot output a binary label for the path.
The second pass of \detector's causality engine tracks the paths on this watchlist and checks whether a new login extends any of these paths;
if so, \detector resubmits each of these newly extended paths for reclassification.

\paragraph{Extending a watchlist path}
For every new login (\newLogin) that occurs, \detector runs both passes of its causality engine.
During this second pass, \detector runs its path inference algorithm to identify all of the watchlist paths that \newLogin extends;
\ie where \detector infers that the final hop of a watchlist path is a causal, inbound login for \newLogin.
If \newLogin extends any watchlist path,
\detector creates a new ``extended path'' ($P_2$) by copying the contents of the watchlist path and
appending \newLogin to $P_2$'s login list.
If the first-pass of \detector's causality engine determined that \newLogin corresponded to a changepoint login, then it updates $P_2$'s attributes accordingly;
otherwise, those values remain the same as the unextended watchlist path.
\detector then takes $P_2$ and submits it to its alert generation algorithm,
which outputs either a binary verdict (benign versus suspicious) or adds $P_2$ to the watchlist.

\paragraph{Pruning the watchlist}
As \detector iterates over the watchlist paths during this second pass,
it removes any path where the final hop occurred over $T$ hours prior to \newLogin, where $T$ reflects the maximum duration of a remote session (\eg 24 hours).
Because the time between the path's last login and \newLogin exceeds the maximum session duration,
the watchlist path could not have caused \newLogin unless the causal user performed a session reconnect or fresh login;
both of these cases will generate a new login event (and corresponding paths) in our data.

\paragraph{Causality Engine's Coverage}
This two-pass approach enables \detector to efficiently trace multi-hop attack paths,
performing only a single-step of backwards tracing during each pass,
while also accurately identifying login paths that involve a (potentially) suspicious use of credentials.
Any path that involves a switch in credentials will contain at least one login that exhibits this switch,
which the first pass of \detector's causality engine will identify.
Each time an attacker switches to a new set of credentials,
\detector will generate an additional set of causal paths that include this credential-switching login in their list of changepoint logins.
For any of these paths, if the attacker does not immediately access a new destination during this credential-switching login,
then \detector's detection algorithms will add the path onto the watchlist.
This step allows the causality engine's second pass to continue tracking these potentially dangerous paths and the destinations they access in the future.

\section{Baseline Evaluation: SAL Details}
\label{appendix:baseline}

In Section~\ref{sec:evaluation:baseline}, we compared the performance of \detector to a prior state-of-the-art detector, \ccsDetector~\cite{ccs2017lmdetecting}.
\ccsDetector requires two-user provided inputs, $N$ the minimum number of days for a benign login edge and $P$ the minimum fraction of machines or users for a benign pattern.
For our comparative evaluation, we performed a grid-search over a range of parameters and selected the best parameters when reporting \ccsDetector's performance: \ie the parameter combination that detected the same number of attacks as \detector with the fewest false positives.
Tables~\ref{table:eval:baseline_full_fp_grid} and~\ref{table:eval:baseline_full_tp_grid} show the total number of false positives and number of attacks detected under a range of parameter combinations that we explored.

\end{document}